# Strain-mediated spin-orbit torque switching for magnetic memory

Qianchang Wang,[1*] John Domann,[2] Guoqiang Yu,[3,4] Anthony Barra,[1] Kang L. Wang,[3,5] Gregory P. Carman[1*]

[1]*Department of Mechanical and Aerospace Engineering, University of California, Los Angeles, CA 90095, USA*
[2]*Department of Biomedical Engineering and Mechanics, Virginia Tech, Blacksburg, VA 24061, USA*
[3]*Department of Electrical and Computer Engineering, University of California, Los Angeles, CA 90095, USA*
[4]*Beijing National Laboratory for Condensed Matter Physics, Institute of Physics, Chinese Academy of Sciences, Beijing 100190, China*
[5]*Department of Physics and Astronomy, University of California, Los Angeles, CA 90095, USA*
[*]*Correspondence and requests for materials should be addressed to Q.W. (email: qianchangwang@ucla.edu) or to G.P.C. (email: carman@seas.ucla.edu)*

## Abstract

Spin-orbit torque (SOT) represents an energy efficient method to control magnetization in magnetic memory devices. However, deterministically switching perpendicular memory bits usually requires the application of an additional bias field for breaking lateral symmetry. Here we present a new approach of field-free deterministic perpendicular switching using a strain-mediated SOT switching method. The strain-induced magnetoelastic anisotropy breaks the lateral symmetry, and the resulting symmetry-breaking is controllable. A finite element model and a macrospin model are used to numerically simulate the strain-mediated SOT switching mechanism. The results show that a relatively small voltage (±0.5 V) along with a modest current ($3.5 \times 10^7$ A/cm$^2$) can produce a 180° perpendicular magnetization reversal. The switching direction ('up' or 'down') is dictated by the voltage polarity (positive or negative) applied to the piezoelectric layer in the magnetoelastic/heavy metal/piezoelectric heterostructure. The switching speed can be as fast as 10 GHz. More importantly, this control mechanism can be potentially implemented in a magnetic random-access memory system with small footprint, high endurance and high tunnel magnetoresistance (TMR) readout ratio.

Deterministic control of magnetism is a key feature for non-volatile magnetic memory devices. One of the most well developed switching mechanisms is spin-transfer torque (STT).[1–3] However, STT-MRAM (magnetic random-access memory) is facing endurance issues because the high current density required for writing damages the thin insulating barrier.[4] For this reason, researchers have begun to investigate spin-orbit torque (SOT) approaches. The SOT switching mechanism offers higher endurance since the write current does not pass through the insulating barrier. Also, SOT is considered to be more energy efficient than STT and theoretically requires lower current density for switching.[5] For magnetic memory devices, memory bits with perpendicular magnetic anisotropy (PMA) are desirable because they have higher thermal stability and smaller footprint compared to in-plane memory bits.[6,7] However, out-of-plane (OOP) switching in memory bits with PMA remains a challenge for SOT devices.

Deterministic OOP switching in SOT devices usually requires an external magnetic bias field,[8–10] but the necessity of the bias field sacrifices energy efficiency and becomes awkward for on-chip application. Field-free deterministic OOP switching has been achieved by breaking the lateral symmetry using the non-uniform oxidation by the wedge shape,[11] tilted anisotropies,[12] an exchange-bias from an adjacent antiferromagnetic layer,[13,14] or a dipole field from an nearby magnetically fixed layer.[15] However, all of these symmetry-breaking methods are non-controllable once the devices are built. To address this issue, researchers have recently developed an SOT device that uses an in-plane electric field to achieve controllable symmetry-breaking.[16] However, this field-free deterministic OOP switching still relies on an induced unidirectional anisotropy, similar to previously listed symmetry-breaking methods. Here, we present a new symmetry-breaking method that uses bidirectional/uniaxial anisotropy instead, which may lead to a new genre of SOT devices with controllable deterministic OOP switching ability.

In particular, this paper demonstrates that strain-induced magnetoelastic anisotropy can feasibly serve as the symmetry-breaking mechanism. The magnetoelastic anisotropy is induced in the magnetoelastic material (e.g., CoFeB) by locally straining an attached piezoelectric layer through externally applied voltage. The strain-induced magnetoelastic anisotropy is uniaxial since the piezo-strains are uniaxial in nature. This form of strain-mediated control represents the most energy efficient way to control magnetism in the nanoscale in current technologies.[17–20]

In this paper, we first show the simulation results of the strain-mediated SOT switching for a representative magnetoelastic/heavy metal/piezoelectric heterostructure. A model that couples micromagnetics, piezoelectricity, elastodynamics, and spin-orbit torque has been used to simulate the switching process. It is demonstrated that the controllable strain-induced magnetoelastic anisotropy can break the lateral symmetry and enable deterministic OOP switching. The switching direction is dictated by the polarity of the applied electric field. Then this fully coupled model is complemented by a macrospin model to plot switching phase diagrams that narrow the design space. The simulation results are followed by the theory explanations. Finally, we present a potential MRAM design based on the strain-mediated SOT method.

**Finite element model setup and results.** Fig. 1a shows the magnetoelastic/heavy metal/piezoelectric heterostructure simulated in the finite element model. The magnetic element is a CoFeB disk with a 50 nm diameter and a 1.5 nm thickness. Underneath the CoFeB disk, there is a canted thin (< 10 nm) heavy metal (e.g., Ta) strip and the SOT current is applied 45° counter-clockwise from the *–y* axis. The SOT current causes spin polarized electrons to accumulate at the magnetoelastic/heavy metal interface. As shown in Fig. 1b, the spin polarization is perpendicular to the applied current as $\hat{\boldsymbol{\sigma}} = (1,1)/\sqrt{2}$, which then exerts spin-orbit torque on the CoFeB magnetic moment.[5,11,21] The heavy metal strip is attached to a 100 nm-thick $Pb[Zr_xTi_{1-x}]O_3$ (PZT) substrate poled along the *z* axis. Two 50 nm × 50 nm electrodes are placed on PZT top surface and are 20 nm from the CoFeB edge along the *y* direction, while the bottom of the PZT is electrically grounded. The assumptions, boundary conditions, and material parameters are described in the Methods section. The CoFeB magnetization is initialized as 'up' (i.e., pointing to the +*z* direction). At t = 0, a −0.5 V voltage is applied to the two top electrodes and a current with density of $5 \times 10^7$ A/cm$^2$ is applied to the heavy metal strip simultaneously. Both voltage/current inputs are removed at t = 2 ns.

Fig. 1c shows the CoFeB disk's volume-averaged biaxial strain $\varepsilon_{bi} = \varepsilon_{yy} - \varepsilon_{xx}$ as a function of time during the application and removal of voltage/current identified by red and blue colors respectively. Due to shear lag effects, the vertical electrical field inside the PZT layer causes anisotropic in-plane strains.[18,22] In this case, a −0.5 V voltage induces volume-averaged biaxial strain $\varepsilon_{bi} \approx 1600\ \mu\varepsilon$ inside the CoFeB disk. It is shown that the biaxial strain rises and oscillates at 1600 $\mu\varepsilon$ within the first 2 ns. Following voltage removal at 2 ns, the strain decays to 0 with small oscillations as the magnetization precesses to its new equilibrium position. The small oscillations are caused by the inverse magnetoelastic effect. The relatively long precession towards the new equilibrium state ($m_z = -1$) is related to the relatively low CoFeB Gilbert damping ($\alpha = 0.01$).

Fig. 1d shows the volume-averaged perpendicular magnetization $m_z$ as a function of time while the inset shows the magnetization trajectory. The voltage/current application and removal are identified as the blue and red portions of the response, respectively. As can be seen, the magnetization stabilizes at 2 ns is down-canted (i.e., $m_z < 0$) before removal of the voltage/current. Therefore, upon removal of both voltage and current (t = 2 ns), the magnetization preferably selects 'down' to its new equilibrium state (i.e., $m_z = -1$). Thus, during this voltage/current application and subsequent removal, the magnetization undergoes 180° OOP switching from 'up' to 'down'.

To better understand the magnetic switching process, Fig. 1e provides magnetic vector diagrams at four representative time points corresponding to the four data points highlighted in Fig. 1d. The arrows indicate in-plane magnetization direction and amplitude while color represents the perpendicular amplitude $m_z$ which varies between −1 and +1. At t = 0, all spins are initialized as 'up', while at t = 0.2 ns the spins rotate to in-plane through 90° switching. The coherent switching observed is attributed to the relatively small CoFeB disk's diameter compared to its single-domain limit.[20,23] After t = 0.2 ns, the magnetization starts to cant down and remain in the –*z* space in the following process. At t = 2 ns, the magnetization stabilizes in a down-canted direction (see Fig. 1d) with $m_z = -0.23$ for this specific voltage/current combination. At t = 12 ns, the spins reach a new equilibrium state ($m_z = -1$). It can be inferred that removing the voltage/current at any time after 0.2 ns will result in deterministic magnetic switching from 'up' to 'down', which corresponds to a 5 GHz writing speed. Increasing voltage or current magnitudes increases writing speed up to ~10 GHz (see Supplementary Note 3 and Supplementary Fig. S3).

For simplicity, in the following results, we only focus on the $m_z$ temporal evolution during the application of voltage/current during t = 0 ~ 2 ns.

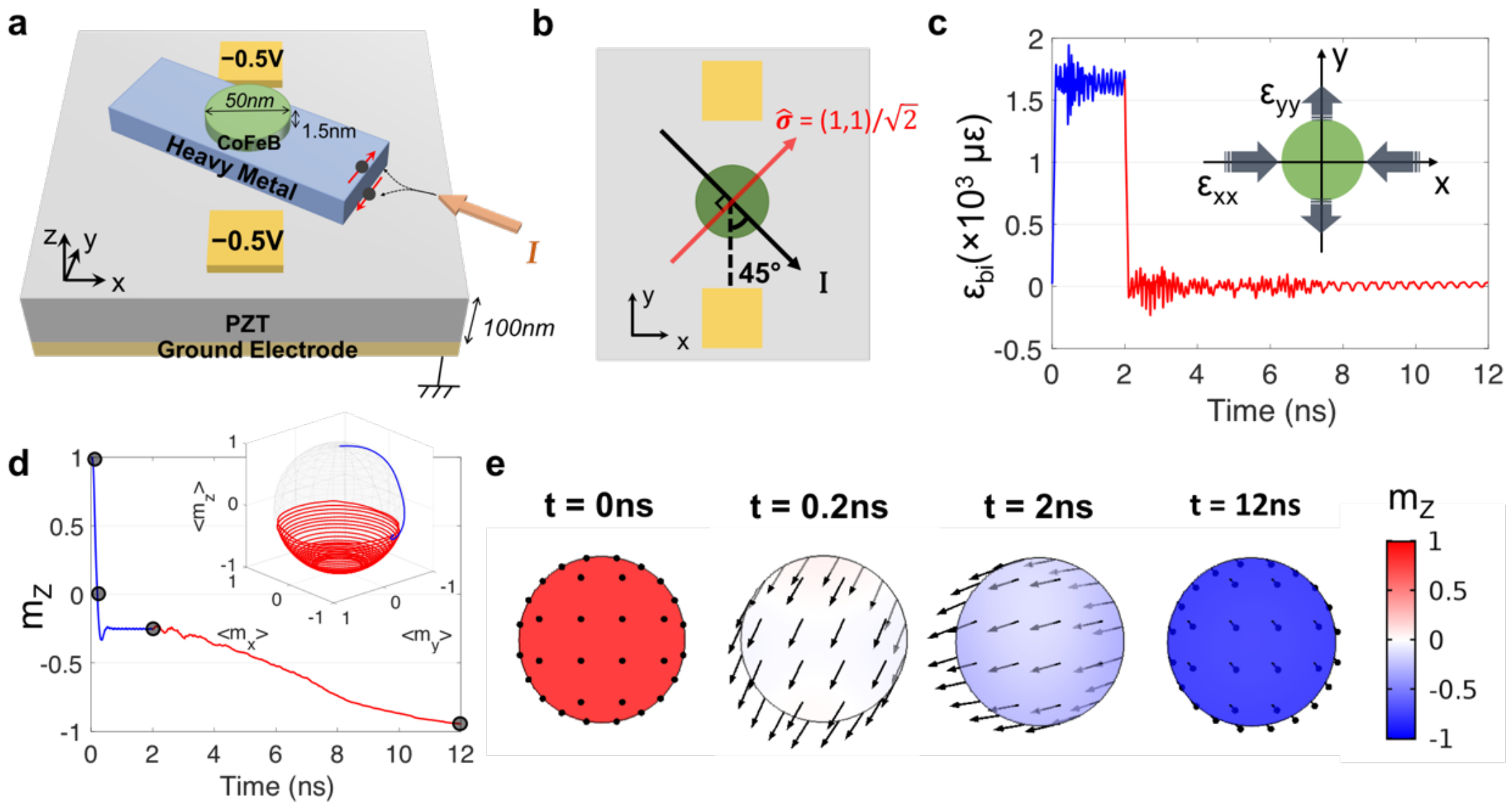

**Figure 1 | Finite element simulation: model setup and results.** (**a**) Schematic illustration of the simulated structure and the direction of applied current. (**b**) The top view of the simulated structure. The spin polarization (red arrow) induced by the SOT current is perpendicular to the current (black arrow). (**c**) Temporal change of biaxial strain $\varepsilon_{bi} = \varepsilon_{yy} - \varepsilon_{xx}$ induced by the −0.5 V applied voltage. In the first 2 ns (blue portion), a −0.5 V voltage is applied to the two top electrodes and simultaneously a $5 \times 10^7$ A/cm$^2$ current density is applied to the heavy metal strip. The red portion represents the evaluation after both voltage and current are removed at t = 2 ns period. The inset figure shows the in-plane principal strain direction. (**d**) Temporal evolution of the perpendicular magnetization in the free CoFeB layer. The four highlighted points are representative time points. The inset figure is the magnetization trajectory. (**e**) Vector diagrams of the magnetization distribution at t = 0, 0.2, 2, and 12 ns, which correspond to the highlighted

points in **d**. The arrows represent the in-plane magnetization amplitude and direction, while the colors represent the perpendicular magnetization $m_z$.

Fig. 2 provides more finite element simulation results to investigate the impact of voltage polarities (positive or negative) and initial states ('up' or 'down') on the final magnetic states. For all cases in Fig. 2, the current density and direction are the same as presented in Fig. 1. Fig. 2a illustrates in-plane principal tension and compression for the −0.5 V case, indicating the corresponding magnetoelastic field is along $\pm y$ axis (not shown here) since CoFeB is a positive magnetostrictive material. Note the field is bidirectional due to the uniaxial nature of the strains. Fig. 2b shows the volume-averaged $m_z$ as a function of time for −0.5 V applied voltage for both 'up' and 'down' initialization. At t = 2 ns, the magnetization is down-canted as $m_z = -0.23$, regardless of the initial state ('up' or 'down'). It can be inferred that the final magnetic states following voltage/current removal are 'down' ($m_z = -1$) for both cases.

If one reverses the current direction in Fig. 2a by 180°, which is not shown here, the final state is the same, i.e., pointing 'down'. This can be explained as follows. If one rotates the *xy* coordinates with respect to *z* axis by 180°, the SOT current reverses direction while the voltage-induced strain remains the same due to its uniaxial nature. This means that the two cases (positive/negative applied current) are physically identical. Therefore, the final states for positive and negative applied current are the same. In contrast, reversing the voltage polarity reverses the final state. As shown in Fig. 2c, for the +0.5 V case, the principal tension and compression directions is reoriented by 90° compared to −0.5 V case shown in Fig. 2a, and the corresponding magnetoelastic field is now along $\pm x$ axis (not shown here). The switching results for the configuration in Fig. 2c are shown in Fig 2d. The magnetic states at t = 2 ns are both up-canted with $m_z = +0.23$ regardless of the initial states. It can be inferred that the final magnetic states following voltage/current removal are 'up' ($m_z = +1$). These results presented in Fig. 2 clearly demonstrate that the switching direction (i.e., 'up' or 'down') is dictated by the voltage polarity (i.e., positive or negative) used to strain the PZT layer. More detailed simulation results can be found in Supplementary Note 2 and Supplementary Fig. S2.

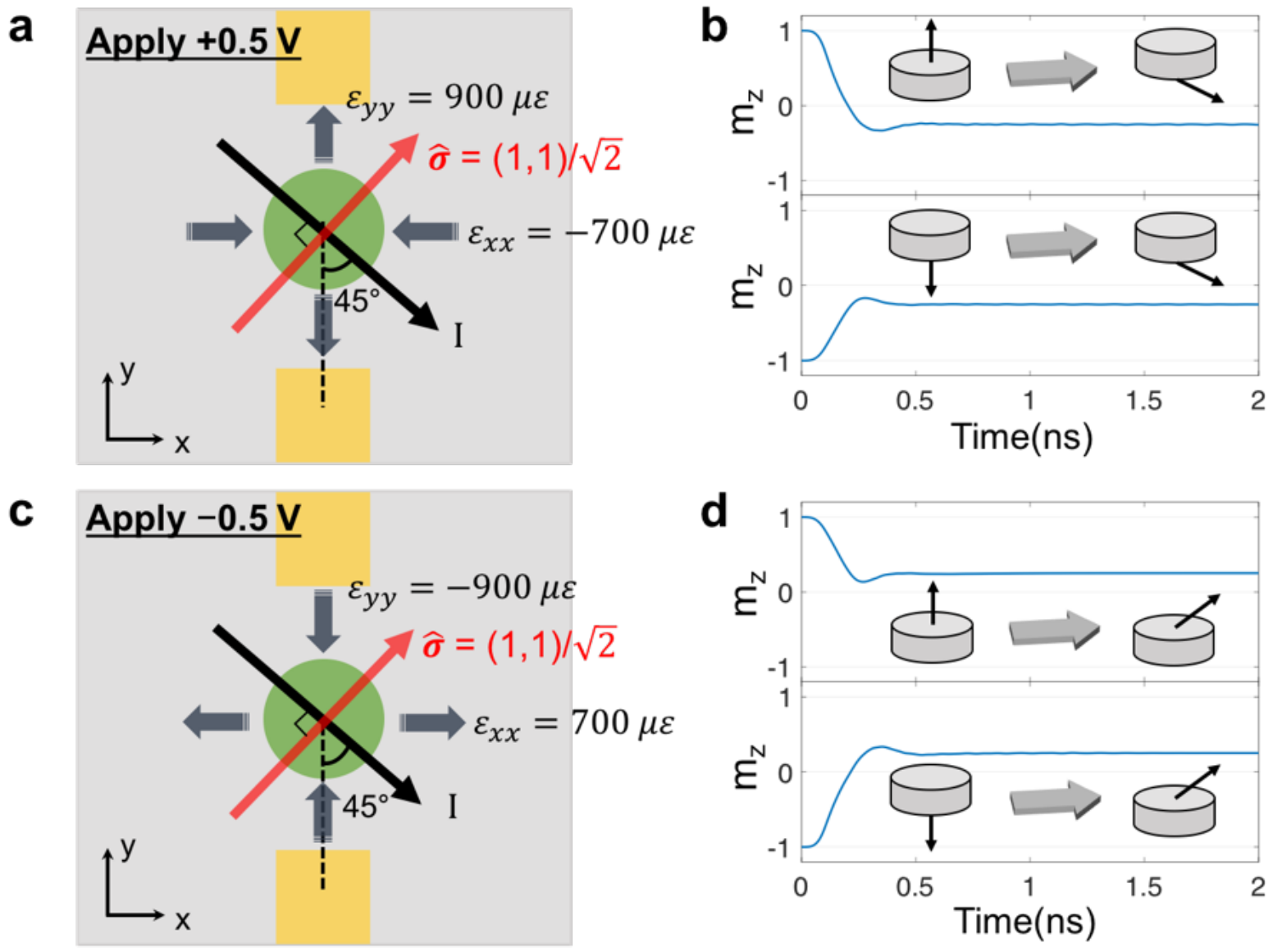

**Figure 2 | Finite element model results of strain-mediated SOT switching for different voltage polarities (positive or negative) and initial states ('up' or 'down').** (**a**) Illustration of strain distribution due to the −0.5 V voltage applied to the two top electrodes. The effective magnetoelastic field $\boldsymbol{H_{ME}}$ for this case is along $\pm y$ axis. The current direction shown in a black arrow is 45° counter-clockwise from $-y$ axis. The induced electron charge polarization is $\hat{\boldsymbol{\sigma}} = (1,1)/\sqrt{2}$. (**b**) For the operating configuration shown in **a**, the final magnetic state is down-canted regardless of the initial states. (**c**) The strain distribution due to the +0.5 V voltage applied to the

two top electrodes. The compression/tension reorient by 90° compared to **a**. The effective magnetoelastic field $\boldsymbol{H_{ME}}$ for this case is along $\pm x$ axis. (**d**) For the operating configuration shown in **c**, the final state is up-canted regardless of the initial states.

**Switching phase diagrams.** In addition to voltage polarity and initial state, there are three additional parameters (i.e., strain amplitude, current density, and relative orientation of strain/current) that influence the strain-mediated SOT switching. Parametric studies using the macrospin model are performed to investigate the influence these three parameters have on the strain-mediated SOT switching, and the results are presented in Fig. 3. In all the cases demonstrated in Fig. 3, the voltage is kept negative and the magnetization is always initialized 'up' (i.e., $m_z = +1$). Details of the macrospin model are described in the Methods section. Supplementary Note 1 and Supplementary Fig. S1 show that the macrospin model provides reasonably accurate results when compared to the fully coupled finite element model. Fig. 3a illustrates the principal strain directions and the definition of the relative orientation θ. From the previous finite element model results, the negative voltage induced strains are tensile along $y$ direction and compressive along $x$ direction. For simplicity, we set $\varepsilon_{xx} = 0$ in the macrospin model and the biaxial strain becomes $\varepsilon_{bi} = \varepsilon_{yy} - \varepsilon_{xx} = \varepsilon_{yy}$. In the parametric studies, the SOT current density is varied from 0 ~ $8 \times 10^7$ A/cm$^2$, the biaxial strain is varied from 0 ~ 4000 μ$\varepsilon$, and θ is varied from 0 ~ 90°.

Fig. 3b illustrates the four types of magnetic state identified at t = 2 ns using the criterion described in the Supplemental Note 6. The type I state represents successful switching with magnetization switching from 'up' to 'down'. The type II state refers to magnetization switching 90° to in-plane where the applied SOT current is sufficiently large to hold the magnetization in-plane. The type III state occurs when the applied voltage and current are both relatively small producing magnetization rotation of less than 90°. Finally, the type IV state represents continued magnetic oscillation, which is beyond the scope of this work and is not discussed in detail.

In the first parametric study, which consists of 2,601 cases (i.e., a 51 × 51 grid), the biaxial strain and the current density are varied while the relative orientation is fixed as θ = 45°. Fig. 3c shows the switching phase diagram, and the four separate regions correspond to the four types of magnetic state shown in Fig. 3b. The successful switching cases (region I) are further examined in Fig. 3d. The $m_z$ amplitude at t = 2 ns is illustrated in color for each case, and the diagram is smoothed using linear interpolation. The switching initiates when the biaxial strain is as low as $\varepsilon_{bi} = \varepsilon_{yy} - \varepsilon_{xx} = 230$ μ$\varepsilon$. As the strain increases, the threshold current (i.e., the minimum current that enables switching) decreases. In other words, a tradeoff exists between the threshold strain and threshold current. In this case, the threshold current reaches a minimum of ~$1 \times 10^7$ A/cm$^2$ at 3000 μ$\varepsilon$ strain amplitude. Further strain increase does not continue to reduce the threshold current because oscillations begin to occur.

The second parametric study also consists of an additional 2,061 cases (i.e., a 51 × 51 grid) with fixed biaxial strain $\varepsilon_{bi} = \varepsilon_{yy} - \varepsilon_{xx} = 1500$ μ$\varepsilon$ while varying θ and current density. Fig. 3e shows the switching phase diagram for this parametric study. Only three types of magnetic states (type I, II, and III) are found, with the magnetic oscillation (type IV) being absent due to the relatively small strains investigated. Fig. 3f further examines all the successful switching cases (region I) and the $m_z$ amplitude at t = 2 ns for each case is represented in the color map. The results show that switching is absent when the current is parallel (θ = 0°) or perpendicular (θ = 90°) to the magnetoelastic field (i.e., $y$ axis). This feature will be explained in the next section using symmetry analysis. For Fig. 3f, both the threshold current and $m_z$ decrease as θ decreases. Smaller threshold currents represent more energy efficient switching process, while lower $m_z$ values represent less reliable switching if thermal fluctuations are included. In other words, for a given voltage, a tradeoff exists between energy efficiency and the reliability of magnetization reversal (or write error). It is also interesting to note that the dashed cut-lines in Figs. 3d and 3f have similar profiles because they both represent switching behaviors as a function of current density for a fixed current direction θ = 45° and a fixed biaxial strain $\varepsilon_{bi} = 1500\ \mu\varepsilon$.

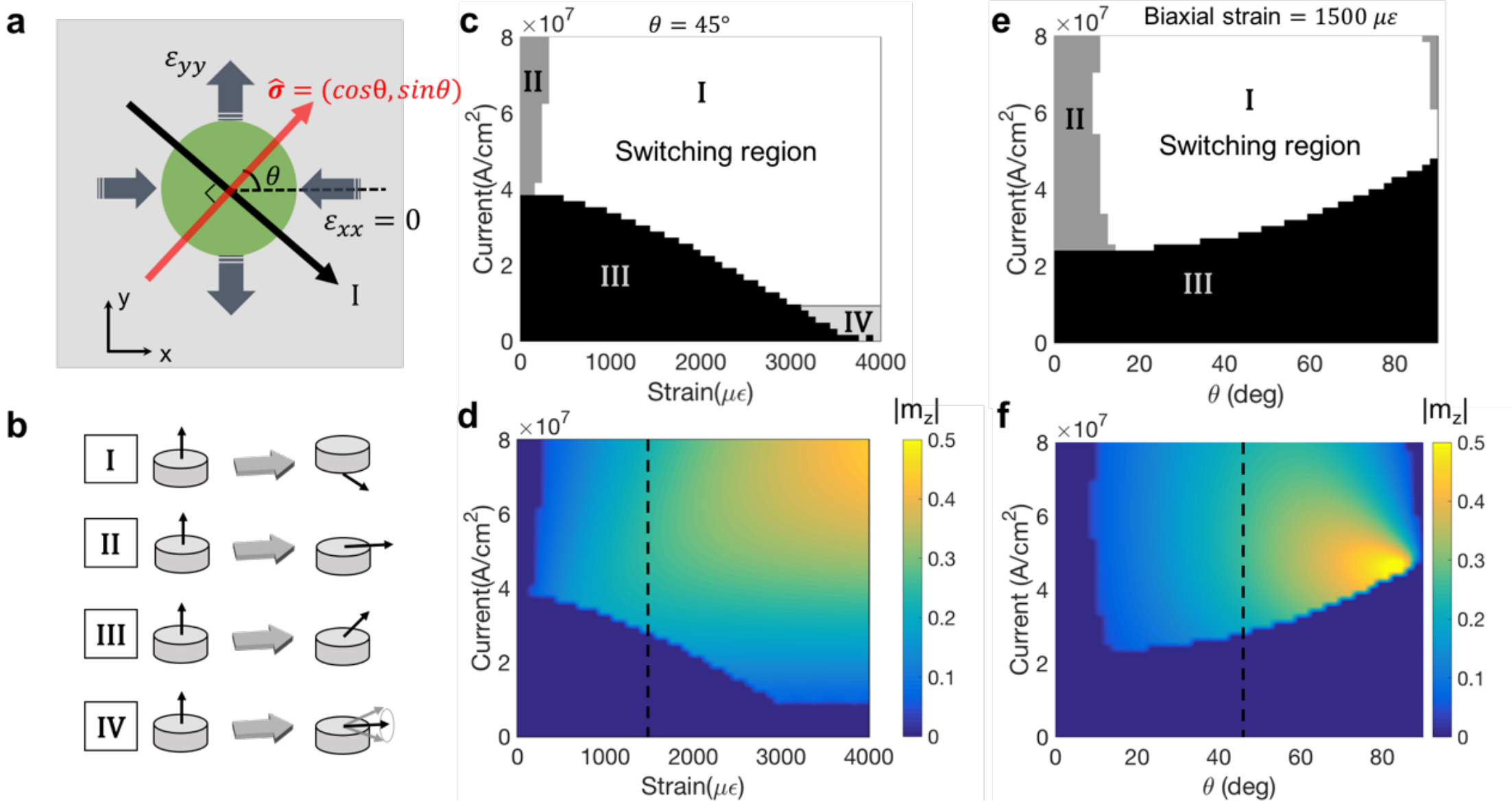

**Figure 3 | Switching phase diagrams drawn from the parametric studies using macrospin model.** (**a**) Definition of θ and in-plane principal strain directions. The biaxial strain is $\varepsilon_{bi} = \varepsilon_{yy} - \varepsilon_{xx} = \varepsilon_{yy}$ assuming $\varepsilon_{xx} = 0$, and the polarization is $\hat{\boldsymbol{\sigma}} = (cos\theta, sin\theta)$. (**b**) Illustrations for different switching behaviors at t = 2 ns found in the simulations, which correspond to region

I, II, III, and IV, respectively in following figures. (**c**) Switching phase diagram for fixed θ = 45° with varying biaxial strain and current density. (**d**) Amplitude of $m_z$ after switching for cases in the region I in **b**. Specifically, the absolute value of $m_z$ is presented here because $m_z < 0$ after the successful switching. (**e**) Switching phase diagram for fixed biaxial strain $\varepsilon_{bi} = 1500$ μ$\varepsilon$ with varying θ and current densities. (**f**) Amplitude of $m_z$ after switching for cases in the region I in **e**. The dashed lines in (**e,f**) both show the switching as a function of current for fixed $\varepsilon_{bi} = 1500$ μ$\varepsilon$ and θ = 45°.

**Symmetry analysis.** To understand the physics behind the deterministic switching, Fig. 4 presents symmetry analysis for three scenarios of strain-mediated SOT switching. In all the configurations shown in Fig. 4, the green sheet represents the magnetic element with PMA, and the grey sheet represents the heavy metal (e.g., Ta). The SOT current *I* is always applied along the –*x*, and the accumulated spin polarization is $\hat{\boldsymbol{\sigma}} = -\hat{y}$. Therefore, the damping-like spin-orbit field is $\boldsymbol{H}_{SO} \propto \boldsymbol{m}\times\hat{\boldsymbol{\sigma}} = \hat{y}\times\boldsymbol{m}$ and the spin-orbit torque is $\boldsymbol{\tau}_{SO} \propto \boldsymbol{H}_{SO}\times\boldsymbol{m}$. The magnetoelastic field is generalized as a uniaxial field $\boldsymbol{H}_{Uni}$ and is represented by a bidirectional arrow.

Figs. 4(a-c) show the symmetry analysis when the $\boldsymbol{H}_{Uni}$ is applied along the direction of the current *I*. In the beginning of the switching process, $\boldsymbol{\tau}_{SO}$ is the only driving torque. Due to the direction of $\boldsymbol{\tau}_{SO}$ relative to the $\boldsymbol{H}_{Uni}$, both directions/branches of $\boldsymbol{H}_{Uni}$ are equally effective. Therefore, applying $\boldsymbol{H}_{Uni}$ along the direction of current is equivalent to applying a bidirectional external bias field $\boldsymbol{H}_b$ along the current. As shown in Fig. 4c, this results in two magnetic stable states in the second and the fourth quadrants in the *xz* plane. Specifically, the branch of $\boldsymbol{H}_{Uni}$ that is parallel to the SOT current prefers an end-state canted 'up', while the branch of $\boldsymbol{H}_{Uni}$ that is anti-

parallel to the SOT current prefers an opposite end-state, i.e. canted 'down'. The dependence of the switching direction on the external field direction agrees with experimental results shown in previous research,[8] where opposite switching behaviors were observed when using Pt as the heavy metal. Note Pt is known to exhibit the opposite SOT switching behavior in contrast to Ta used in our simulation.[24] In conclusion, when the uniaxial field is applied along the SOT current direction, the symmetry is not broken, and deterministic switching is not produced.

Figs. 4(d-f) show the symmetry analysis when $\boldsymbol{H_{Uni}}$ is applied 45° clockwise relative to current $I$ in the $xy$ plane. One can easily recognize the configurations in Figs. 4d and 4e are the same as Fig. 2a, where −0.5 V voltage is applied to the PZT top electrodes. It is shown here that wherever the magnetization initializes, either 'up' (Fig. 4d) or 'down' (Fig. 4e), only one branch of $\boldsymbol{H_{Uni}}$ is effective in the switching process due to the direction of $\boldsymbol{\tau_{SO}}$. Interestingly, the same $\boldsymbol{H_{Uni}}$ branch is effective for both initial magnetic states in Figs. 4d and 4e. More specifically, because $\boldsymbol{\tau_{SO}}$ is along positive $+y$ axis in both Figs. 4d and 4e, the $\boldsymbol{H_{Uni}}$ branch with positive $y$ component is effective. The effective branch of $\boldsymbol{H_{Uni}}$ is shown in red, while the dummy branch of $\boldsymbol{H_{Uni}}$ is shown in light pink. Note the field component that is perpendicular to the SOT current does not contribute to the symmetry breaking. Because the projection of $\boldsymbol{H_{Uni}}$ onto the SOT current is anti-parallel to the current, it is equivalent to applying an external bias field anti-parallel to the current, as shown in Fig. 4f. Therefore, the symmetry is broken and only a down-canted state in the fourth quadrant is allowed, while the states in other quadrants are unstable/prohibited. This agrees with the simulation results presented in Fig. 2b, which showed that final states are always down-canted regardless of initial states.

Figs. 4(g-i) show the symmetry analysis when $\boldsymbol{H_{Uni}}$ is applied 45° counter-clockwise relative to the current *I* in *xy* plane. This configuration is the same as in Fig. 2c, where +0.5 V voltage is applied to the PZT top electrodes. Similar to previous situation shown in Fig. 4(c-e), only one branch of $\boldsymbol{H_{Uni}}$ is effective during the switching process depending on the direction of $\boldsymbol{\tau_{SO}}$. However, the projection of the effective $\boldsymbol{H_{Uni}}$ onto the SOT current is now parallel to the current. Therefore, the effective $\boldsymbol{H_{Uni}}$ is equivalent to applying an external bias field parallel to the current, as shown in Fig. 4i. The symmetry is also broken and the up-canted state is selected. This agrees with the simulation results presented in Fig. 2d, i.e., final states are always up-canted regardless of initial states. In complement to the symmetry analysis presented in Fig. 4, two equivalent symmetry analysis methods are provided in Supplemental Information, including the mirror symmetry analysis (Note 4) and mathematical derivations using divergence arguments (Note 5).

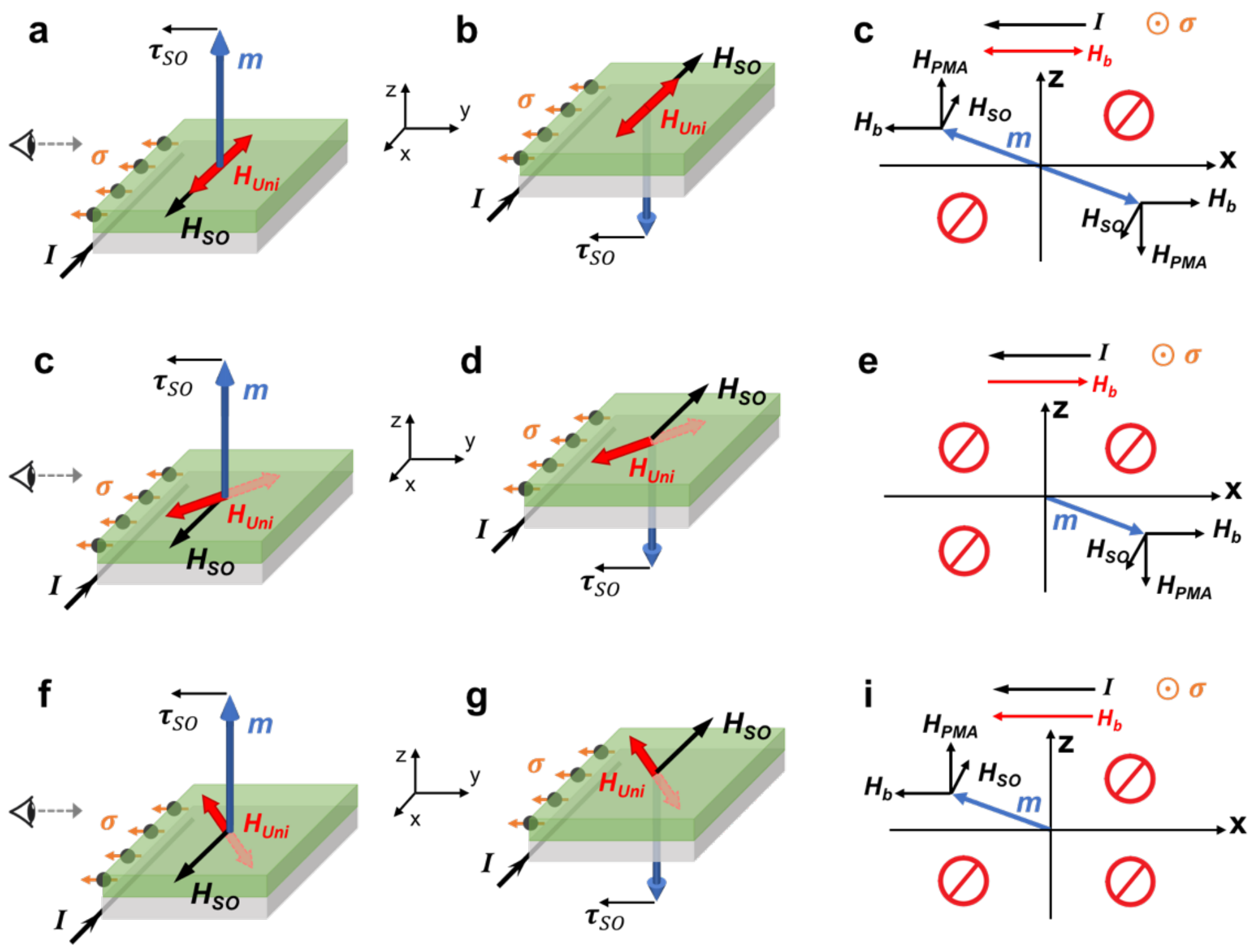

**Figure 4 | Symmetry analysis of strain-mediated SOT switching.** The strain-induced magnetoelastic field is generalized as $\boldsymbol{H_{Uni}}$. (**a-c**) Symmetry analysis for the scenario in which $\boldsymbol{H_{Uni}}$ is parallel to the SOT current. Due to the uniaxial nature, $\boldsymbol{H_{Uni}}$ is equivalent to a bidirectional external bias field $\boldsymbol{H_b}$. The symmetry is not broken, resulting in two equally likely equilibrium end-state configurations. This means that the switching, in this case, is not deterministic. (**d-f**) Symmetry analysis for the scenario in which $\boldsymbol{H_{Uni}}$ is applied 45° clockwise relative to current $I$ in the $xy$ plane. Due of the presence of spin-orbit torque $\boldsymbol{\tau_{SO}}$, only one branch of the $\boldsymbol{H_{Uni}}$ is effective. This is equivalent to applying a bias field that is anti-parallel to the SOT current, and only the down-canted state is favorable. This corresponds to the situation in Fig. 2(**a,b**). (**g-i**) Symmetry analysis for the scenario in which $\boldsymbol{H_{Uni}}$ is applied 45° counter-clockwise relative to current $I$ in the $xy$ plane. Also, due of the presence of spin-orbit torque $\boldsymbol{\tau_{SO}}$, only one

branch of the $\boldsymbol{H_{Uni}}$ is effective. This is equivalent to applying a bias field that is parallel to the SOT current, and only the up-canted state is favorable. This corresponds to the situation in Fig. 2(**c,d**).

**Architecture of MeSOT-RAM.** Based on the strain-mediated SOT switching mechanism demonstrated above, we are able to design a magnetic memory system. Fig. 5 shows the proposed Magnetoelastic Spin-Orbit Torque Random Accessed Memory (MeSOT-RAM). Fig. 5a illustrates a representative 2 × 2 memory array architecture. Each memory bit has a perpendicular magnetic tunnel junction (pMTJ) stack whose free layer and reference layer are both CoFeB with PMA. As a result, the memory device has relatively small footprint hence high storage capacity. All bits are connected in rows by bitlines (BL) and in columns by platelines (PL), and the BL/PL are designed to tilt from each other. Fig. 5b shows the suggested materials in the MeSOT-RAM. Writing a bit requires concurrently applying voltage/current to the PL/BL, respectively. Applying +0.5V/−0.5V and a SOT current results in the magnetization of free layer switching deterministically up/down, respectively (i.e., writing '1' or '0'). Random access is feasible since neither voltage nor current is sufficient to switch the pMTJ by itself. Readout from this design is achieved by applying voltage across the wordline (WL) and BL and measuring the pMTJ's magnetoresistance, similar to reading in conventional MRAMs. High TMR readout ratio is promised because the symmetry-breaking by magnetoelastic anisotropy is in-situ controllable. In addition, the separate writing and reading pathways provide relatively high endurance and low write error rate.

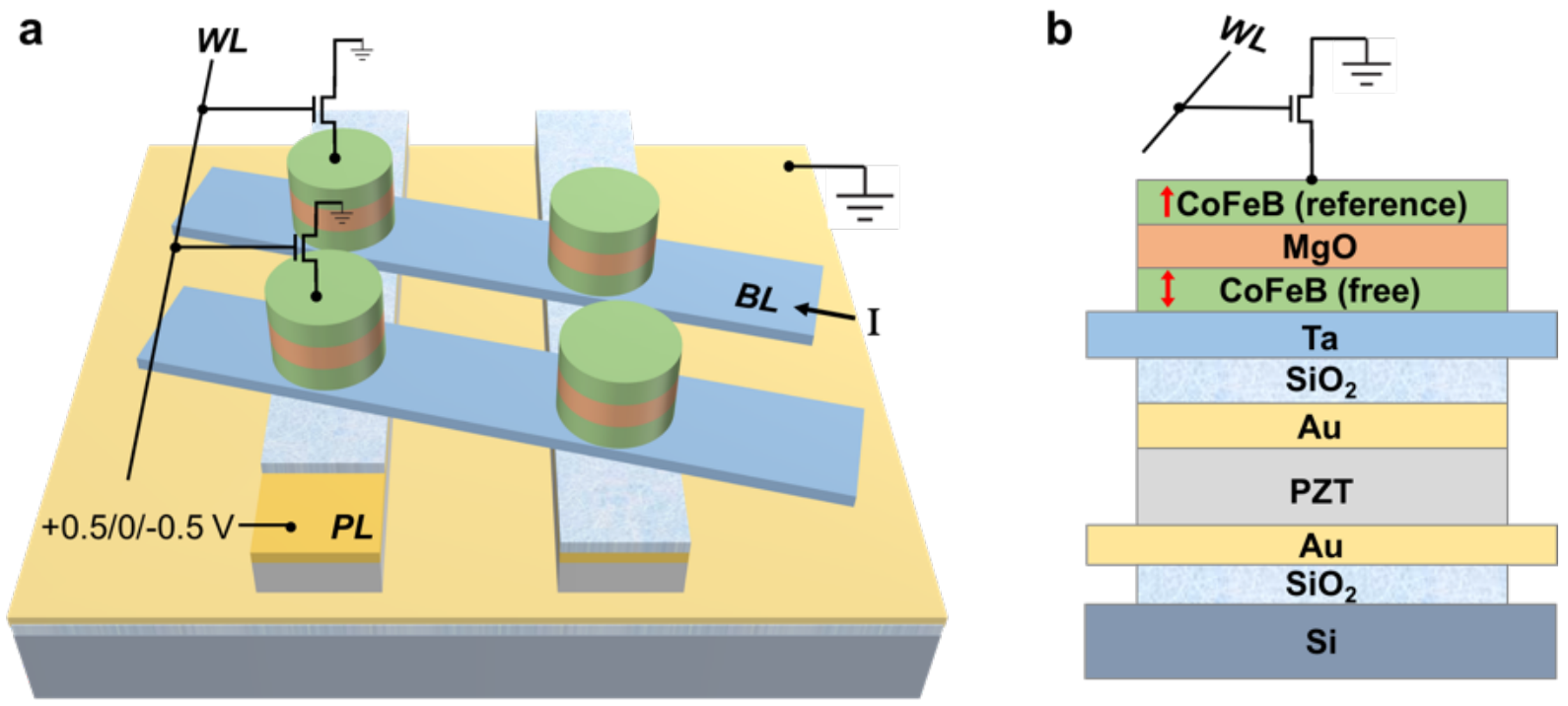

**Figure 5 | Schematics of MeSOT-RAM based on strain-mediated SOT switching mechanism.** (**a**) Architecture of a 2 × 2 memory array of MeSOT-RAM. Each memory bit consists of a pMTJ and a transistor. Writing is achieved by concurrently applying voltage/current to the plateline/bitline (PL/BL), respectively. Reading is achieved by applying voltage between BL and wordline (WL) and measuring the magnetoresistance across the pMTJ stack. (**b**) Illustration of a single memory bit and the suggested material stack. The red arrows in CoFeB layers represent the possible magnetization. The reference CoFeB layer has fixed magnetization pinned by antiferromagnetic layer (not shown). The free CoFeB's magnetization can be switched through the strain-mediated SOT switching mechanism and the 'up' and 'down' magnetization states are encoded as information '1' or '0'.

In conclusion, a finite element model and a macrospin model are used to simulate strain-mediated SOT switching of a nanomagnetic structure with PMA. The switching is field-free, fast, and deterministic. The final magnetic state depends on the polarity of the applied voltage, but does not depend on the initial state. The lowest threshold current required for the switching is as small as $1 \times 10^7$ A/cm$^2$ with the lowest threshold biaxial strain required for switching is 230 $\mu\varepsilon$. Optimizing the structure geometries and materials may further decrease the threshold current and threshold strain thus increase energy efficiency. More importantly, this symmetry-breaking mechanism is universally applicable to other uniaxial anisotropy, such as voltage-controlled magnetic anisotropy (VCMA). Using the uniaxial anisotropy to break the in-plane symmetry opens a new genre of field-free deterministic OOP switching in SOT devices and paves the way for next-generation non-volatile memory.

## Methods

**Finite element model**. A fully coupled micromagnetic and elastodynamic finite element model is developed using multiphysics software to simulate the strain-mediated SOT switching. The precessional magnetic dynamics are governed by the Landau-Lifshitz-Gilbert (LLG) equation with SOT terms:[21,25,26]

$$\frac{\partial \boldsymbol{m}}{\partial t} = -\mu_0 \gamma \left(\boldsymbol{m} \times \boldsymbol{H}_{eff}\right) + \alpha \left(\boldsymbol{m} \times \frac{\partial \boldsymbol{m}}{\partial t}\right) - \frac{\gamma \hbar}{2e} \frac{J_C}{M_S t_F} \xi_{DL} \boldsymbol{m} \times (\boldsymbol{m} \times \widehat{\boldsymbol{\sigma}}) \tag{1}$$

where $\boldsymbol{m}$ is the normalized magnetization, $\mu_0$ the vacuum permittivity, $\gamma$ the gyromagnetic ratio, $\hbar$ the reduced Planck constant, $e$ the elementary charge, $J_C$ the current density, $\widehat{\boldsymbol{\sigma}}$ the polarized spin accumulation, $\alpha$ the Gilbert damping factor, $t_F$ the thickness of the free magnetic layer, $M_S$ the saturation magnetization, and $\xi_{DL}$ is the damping-like spin Hall angle. Note the field-like term is not considered in the calculation since it is believed to have no deterministic effect on the magnetization switching.[27] Additionally, $\boldsymbol{H}_{eff}$ is the effective field and consists of four components: $\boldsymbol{H}_{eff} = \boldsymbol{H}_{ex} + \boldsymbol{H}_{Demag} + \boldsymbol{H}_{PMA} + \boldsymbol{H}_{ME}$, where $\boldsymbol{H}_{ex}$ is the exchange field, $\boldsymbol{H}_{Demag}$ the demagnetization field, $\boldsymbol{H}_{PMA}$ the effective PMA field, and $\boldsymbol{H}_{ME}$ the magnetoelastic field. The PMA field is generalized in the following equation using a phenomenological PMA coefficient $K_{PMA}$: $\boldsymbol{H}_{PMA} = -\frac{2}{\mu_0 M_S} K_{PMA} m_z \hat{\boldsymbol{z}}$.[20,28] Specifically, for materials where the PMA originates from interfacial effects (e.g., CoFeB), the PMA coefficient is written as $K_{PMA} = -K_i/t_F$, where $K_i =$ 1.3 mJ/m$^2$ for CoFeB and $t_F$ is the free layer thickness.[7] The magnetoelastic field $\boldsymbol{H}_{ME}$ is obtained by taking the derivative of magnetoelastic energy density with respect to the magnetization $\boldsymbol{m}$:[29]

$$\boldsymbol{H}_{ME}(\boldsymbol{m},\boldsymbol{\varepsilon}) = -\frac{1}{\mu_0 M_S}\frac{\partial E_{ME}}{\partial \boldsymbol{m}} = -\frac{1}{\mu_0 M_S}\frac{\partial}{\partial \boldsymbol{m}}\{B_1[\varepsilon_{xx}\left(m_x^2-\frac{1}{3}\right)+\varepsilon_{yy}\left(m_y^2-\frac{1}{3}\right)$$
$$+\varepsilon_{zz}\left(m_z^2-\frac{1}{3}\right)] + 2B_2(\varepsilon_{xy}m_x m_y + \varepsilon_{yz}m_y m_z + \varepsilon_{zx}m_z m_x)\} \tag{2}$$

where $m_x$, $m_y$ and $m_z$ are components of normalized magnetization along *x, y* and *z* axis, $B_1$ and $B_2$ are first and second order magnetoelastic coupling coefficients. The magnetic material is assumed to be polycrystalline allowing magnetocrystalline anisotropy to be neglected. The formula for calculating demagnetization and exchange field is given by Liang *et al.*[22] thus not repeated here. Thermal fluctuations are neglected in all calculations.

Assuming linear elasticity and piezoelectricity, the behavior of the piezoelectric thin film follows:

$$\boldsymbol{\varepsilon}_p = s_E : \boldsymbol{\sigma} + d^t \cdot \boldsymbol{E} \tag{3}$$

$$\boldsymbol{D} = d : \boldsymbol{\sigma} + e_\sigma \cdot \boldsymbol{E} \tag{4}$$

where $\boldsymbol{\varepsilon}_p$ is strain contribution from piezoelectric effect, $\boldsymbol{\sigma}$ is stress, $\boldsymbol{D}$ is electric displacement, $\boldsymbol{E}$ is electric field, $s_E$ is the piezoelectric compliance matrix under constant electric field, $d$ and $d^t$ are the piezoelectric coupling matrix and its transpose, and $e_\sigma$ is electric permittivity matrix measured under constant stress. The total strain is expressed as $\boldsymbol{\varepsilon} = \boldsymbol{\varepsilon}_p + \boldsymbol{\varepsilon}^{\boldsymbol{m}}$, where $\boldsymbol{\varepsilon}_{ij}^{\boldsymbol{m}} = 1.5\,\lambda_s(m_i m_j - \delta_{ij}/3)$ represents the strain contribution due to isotropic magnetostriction, $\delta_{ij}$ is Kronecker function, and $\lambda_s$ is saturation magnetostriction coefficient.[29]

The magnetization and displacement are simultaneously computed by numerically solving the coupled LLG equation, electrostatic equation, and elastodynamic equation in time domain using finite element method. The strain transferred from the piezoelectric layer to magnetic layer influences the magnetization through $\boldsymbol{H}_{ME}$, and conversely the magnetization change influences the strain through $\boldsymbol{\varepsilon}^{\boldsymbol{m}}$. In other word, the mechanics and magnetics in the finite element model are fully coupled and the coupling is bidirectional.

The piezoelectric substrate simulated in this work is Pb(Zr,Ti)$O_3$ (PZT) poled along the $\hat{z}$ direction. The bottom surface of the PZT is mechanically fixed and electrically grounded while the top surface is traction free. The CoFeB material parameters are: $\alpha = 0.01$, $M_s = 1.2\times10^6$ A/m, $B_1 = B_2 = -2.77\times10^7$ N/m$^2$, exchange stiffness $A_{ex} = 2\times10^{-11}$ J/m (used in $\boldsymbol{H}_{ex}$), $\lambda_s =$ 150 ppm, Young's modulus $E$ = 160 GPa, density $\rho = 7700$ kg/m$^3$, Poisson's ratio $\nu = 0.3$ and spin Hall angle $\xi_{DL}$ = 1.[7,30–32] Note the heavy metal is sufficiently thin that the impact of the heavy metal on the CoFeB strain distribution is neglected in the mechanical analysis. Also, the voltage (~1 mV) required to generate the SOT current is trivially small compared to the PZT (~0.5 V) and is thus neglected in electrostatics calculation. To avoid divergence, the magnetization is always initialized from a slightly canted direction ($m_x$, $m_y$, $m_z$) = (0.1, 0.1, ±0.99). And a 1 ns temporal evolution is performed before any application of current/voltage to allow the magnetization to settle down.

**Macrospin model in MATLAB**. A simplified macrospin model is developed for faster simulation. In the macrospin model, a single spin is used to represent the magnetization of the entire disk. The strain is assumed uniform and constant throughout the magnetic element and does not vary spatially as in the finite element model. Although the macrospin model does not take into consideration the non-uniform strain distribution or the converse magnetoelastic coupling, it is beneficial because the computation time for each temporal evolution is only ~1/1000 of the time consumed in the fully coupled model. The LLG equation is the same as used in the finite element model. However, the total effective field for the macrospin model only consists of three components: $\boldsymbol{H}_{eff} = \boldsymbol{H}_{Demag} + \boldsymbol{H}_{PMA} + \boldsymbol{H}_{ME}$. The exchange field $\boldsymbol{H}_{ex}$ is absent due to the assumption of uniform magnetization, which is reasonable for small structures such as the 50 nm diameter CoFeB disk modeled in this work. In addition, the demagnetization field is simplified to $\boldsymbol{H}_{Demag} = -\boldsymbol{N}M_s\boldsymbol{m}$, where $\boldsymbol{N}$ is the demagnetization tensor for an oblate spheroid.[33] The PMA and magnetoelastic effective fields, as well as all material parameters used in macrospin model are the same with those used in the finite element model. Similar to the finite element model, to avoid singularity solution, the magnetization is always initialized from a slightly canted direction ($m_x$, $m_y$, $m_z$) = (0.1, 0.1, ±0.99) in the macrospin model.

## Acknowledgements

This work was supported by NSF Nanosystems Engineering Research Center for Translational Applications of Nanoscale Multiferroic Systems (TANMS) Cooperative Agreement Award EEC-1160504 and CNSI/HP Nano-Enabled Memory Systems funding mechanism.

## Author contributions

Q.W. conceived and performed the simulations. J.D. provided the macrospin modeling framework. G.Y. mediated in symmetry analysis. K.L.W. and G.P.C. directed the work. A.B. discussed the results. Q.W. and G.P.C. wrote the manuscript. All authors commented the manuscript.

**Reference:**

1. Huai, Y. Spin-Transfer Torque MRAM ( STT-MRAM ): Challenges and Prospects. *AAPPS Bull.* **18,** 33–40 (2008).
2. Diao, Z. *et al.* Spin-transfer torque switching in magnetic tunnel junctions and spin-transfer torque random access memory. *J. Phys. Condens. Matter* **19,** 165209 (2007).
3. Kishi, T. *et al.* Lower-current and Fast switching of A Perpendicular TMR for High Speed and High density Spin-Transfer-Torque MRAM. *IEEE Int.* 1–4 (2008).
4. Cubukcu, M. *et al.* Spin-orbit torque magnetization switching of a three-terminal perpendicular magnetic tunnel junction Spin-orbit torque magnetization switching of a three-terminal perpendicular magnetic tunnel junction. **104,** 42406 (2014).
5. Liu, L., Lee, O. J., Gudmundsen, T. J., Ralph, D. C. & Buhrman, R. A. Current-induced switching of perpendicularly magnetized magnetic layers using spin torque from the spin hall effect. *Phys. Rev. Lett.* **109,** 96602 (2012).
6. Nishimura, N. *et al.* Magnetic tunnel junction device with perpendicular magnetization films for high-density magnetic random access memory. *J. Appl. Phys.* **91,** 5246–5249 (2002).
7. Ikeda, S. *et al.* A perpendicular-anisotropy CoFeB-MgO magnetic tunnel junction. *Nat. Mater.* **9,** 721–724 (2010).
8. Miron, I. M. *et al.* Perpendicular switching of a single ferromagnetic layer induced by in-plane current injection. *Nature* **476,** 189–193 (2011).
9. Qiu, X. *et al.* Spin–orbit-torque engineering via oxygen manipulation. *Nat. Nanotechnol.* **10,** 333–338 (2015).
10. Fan, Y. *et al.* Magnetization switching through giant spin-orbit torque in a magnetically doped topological insulator heterostructure. *Nat. Mater.* **13,** 699 (2014).
11. Yu, G. *et al.* Switching of perpendicular magnetization by spin-orbit torques in the absence of external magnetic fields. *Nat. Nanotechnol.* **9,** 548 (2014).
12. You, L. *et al.* Switching of perpendicularly polarized nanomagnets with spin orbit torque without an external magnetic field by engineering a tilted anisotropy. *Proc. Natl. Acad. Sci.* **112,** 10310–10315 (2015).
13. Oh, Y.-W. *et al.* Field-free switching of perpendicular magnetization through spin-orbit torque in antiferromagnet/ferromagnet/oxide structures. *Nat. Nanotechnol.* **11,** 878–884 (2016).
14. van den Brink, A. *et al.* Field-free magnetization reversal by spin-Hall effect and exchange bias. *Nat. Commun.* **7,** 10854 (2016).
15. Chen, J., Zhang, D., Zhao, Z., Li, M. & Wang, J. Field-free spin-orbit torque switching of composite perpendicular CoFeB / Gd / CoFeB layers utilized for three-terminal magnetic tunnel junctions. *Appl. Phys. Lett.* **111,** 12402 (2017).
16. Cai, K. *et al.* Electric field control of deterministic current-induced magnetization switching in a hybrid ferromagnetic/ferroelectric structure. *Nat. Mater.* **16,** 712 (2017).
17. Wang, K. L., Alzate, J. G. & Khalili Amiri, P. Low-power non-volatile spintronic memory: STT-RAM and beyond. *J. Phys. D. Appl. Phys.* **46,** 74003 (2013).
18. Cui, J. *et al.* A method to control magnetism in individual strain-mediated magnetoelectric islands. *Appl. Phys. Lett.* **103,** 232905 (2013).
19. Zhao, Z. *et al.* Giant voltage manipulation of MgO-based magnetic tunnel junctions via localized anisotropic strain: A potential pathway to ultra-energy-efficient memory technology. *Appl. Phys. Lett.* **109,** 92403 (2016).

20. Wang, Q. *et al.* Strain-mediated 180 switching in CoFeB and Terfenol-D nanodots with perpendicular magnetic anisotropy. *Appl. Phys. Lett.* **110,** 102903 (2017).
21. Fukami, S., Anekawa, T., Zhang, C. & Ohno, H. A spin–orbit torque switching scheme with collinear magnetic easy axis and current configuration. *Nat. Nanotechnol.* **11,** 621–625 (2016).
22. Liang, C. Y. *et al.* Electrical control of a single magnetoelastic domain structure on a clamped piezoelectric thin film - Analysis. *J. Appl. Phys.* **116,** 123909 (2014).
23. Al-Rashid, M. M., Bandyopadhyay, S. & Atulasimha, J. Dynamic Error in Strain-Induced Magnetization Reversal of Nanomagnets Due to Incoherent Switching and Formation of Metastable States: A Size-Dependent Study. *IEEE Trans. Electron Devices* **63,** 3307–3313 (2016).
24. Liu, L. *et al.* Spin-Torque Switching with the Giant Spin Hall Effect of Tantalum. *Science* **336,** 555–558 (2012).
25. Park, J., Rowlands, G. E., Lee, O. J., Ralph, D. C. & Buhrman, R. A. Macrospin modeling of sub-ns pulse switching of perpendicularly magnetized free layer via spin-orbit torques for cryogenic memory applications. *Appl. Phys. Lett.* **105,** 102404 (2014).
26. Finocchio, G., Carpentieri, M., Martinez, E. & Azzerboni, B. Switching of a single ferromagnetic layer driven by spin Hall effect. *Appl. Phys. Lett.* **102,** (2013).
27. Wang, Z., Zhao, W., Deng, E., Klein, J. & Chappert, C. Perpendicular-anisotropy magnetic tunnel junction switched by spin-Hall-assisted spin-transfer torque. *J. Phys. D. Appl. Phys.* **48,** 65001 (2015).
28. Li, X. *et al.* Strain-mediated 180° perpendicular magnetization switching of a single domain multiferroic structure. *J. Appl. Phys.* **118,** 14101 (2015).
29. O'handley, R. C. *Modern magnetic materials*. (Wiley, 2000).
30. Kim, K. S. *et al.* Magnetic and Structural Properties of FeCoB Thin Films. *J. Koerean Phys. Soc.* **54,** 886–890 (2009).
31. Sato, H. *et al.* CoFeB Thickness Dependence of Thermal Stability Factor in CoFeB/MgO Perpendicular Magnetic Tunnel Junctions. *IEEE Magn. Lett.* **3,** 3000204 (2012).
32. Wang, D., Nordman, C., Qian, Z., Daughton, J. M. & Myers, J. Magnetostriction effect of amorphous CoFeB thin films and application in spin-dependent tunnel junctions. *J. Appl. Phys.* **97,** 10C906 (2005).
33. Fratta, G. Di. The Newtonian potential and the demagnetizing factors of the general ellipsoid. *Proc. R. Soc. A* **472,** 20160197 (2016).

SUPPLEMENTARY INFORMATION

# Strain-mediated spin-orbit torque switching for magnetic memory

Qianchang Wang,[1*] John Domann,[2] Guoqiang Yu,[3,4] Anthony Barra,[1] Kang L. Wang,[3,5] Gregory P. Carman[1*]

[1]*Department of Mechanical and Aerospace Engineering, University of California, Los Angeles, CA 90095, USA*
[2]*Department of Biomedical Engineering and Mechanics, Virginia Tech, Blacksburg, VA 24061, USA*
[3]*Department of Electrical and Computer Engineering, University of California, Los Angeles, CA 90095, USA*
[4]*Beijing National Laboratory for Condensed Matter Physics, Institute of Physics, Chinese Academy of Sciences, Beijing 100190, China*
[5]*Department of Physics and Astronomy, University of California, Los Angeles, CA 90095, USA*
[*]*Correspondence and requests for materials should be addressed to Q.W. (email: qianchangwang@ucla.edu) or to G.P.C. (email: carman@seas.ucla.edu)*

**Table of Contents:**

**Supplementary Note 1. Comparison of the finite element model and the macrospin model**

Supplementary Fig. S1 compares the simulation results from finite element model and macrospin model for the structure shown in Fig. 1a (see main text of the paper). The two simulations have same inputs: strain $\varepsilon_{yy} = 900$ με and $\varepsilon_{xx} = -700$ με, current density 5 × $10^7$ A/$cm^2$, and angle θ = 45°, θ is defined in Supplementary Fig. S1. The shadowed region (t = 0 ~ 2 ns) represents the time in which the strain and current are applied. In this region, the two models produce very similar magnetization reorientation results. The main reason for this agreement is the magnetoelastic field $\boldsymbol{H}_{ME}$ dominates to orientation and both models have the same form for $\boldsymbol{H}_{ME}$. However, after the strain and current are removed (2 ~ 12ns period), the two lines become slightly different. This difference is associated with the internal rise in exchange anisotropy (due to slight variations in **m**) which the finite element model considers but is neglected in the macrospin model. Regardless, Supplementary Fig. S1 shows the macrospin model is sufficiently accurate to simulate the strain-mediated SOT switching.

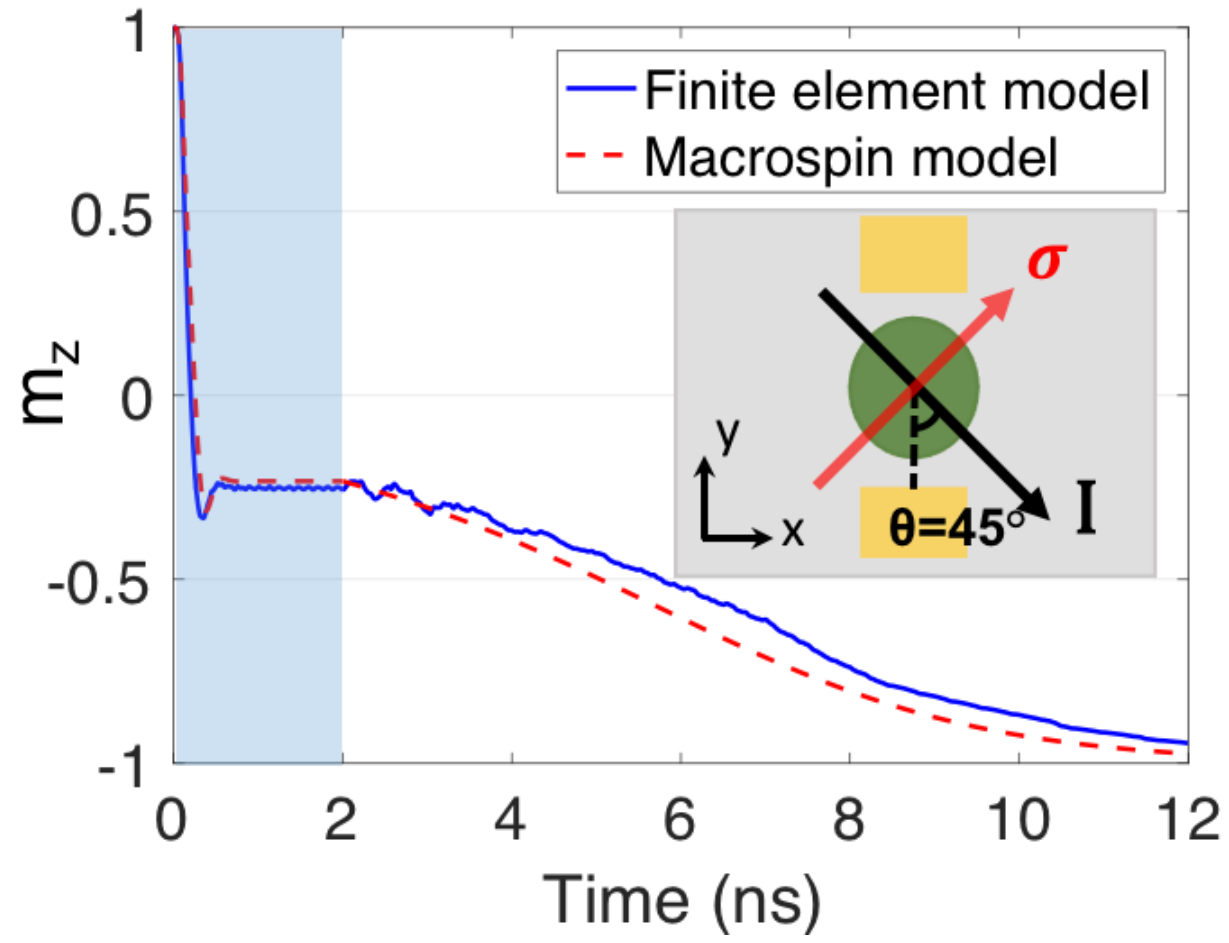

**Supplementary Figure S1.** Simulation results of finite element model and macrospin model for the same structure shown in Fig. 1a in the main text.

**Supplementary Note 2. Impact of initial states**

Supplementary Fig. S2 shows macrospin model results for different initial magnetic states. For all cases, a voltage of ±0.5V and a current density 5 × $10^7$ A/$cm^2$ is applied for the geometry shown in Fig. 1a in the main text. The initial state $m_z$ is varied from −0.9 to +0.9. As shown in Supplementary Fig. S2(a), the final state for +0.5V applied voltage is always canting upward ($m_z = +0.23$) regardless of the initial states. And Supplementary Fig. S2(b) indicates that the final state for −0.5V applied voltage is always canting downward ($m_z = -0.23$) regardless of the initial states. In summary, the initial state has no impact on the final state. This is in complementary to Supplementary Fig. 2 in the main text to prove that the final state is not dependent on the initial state.

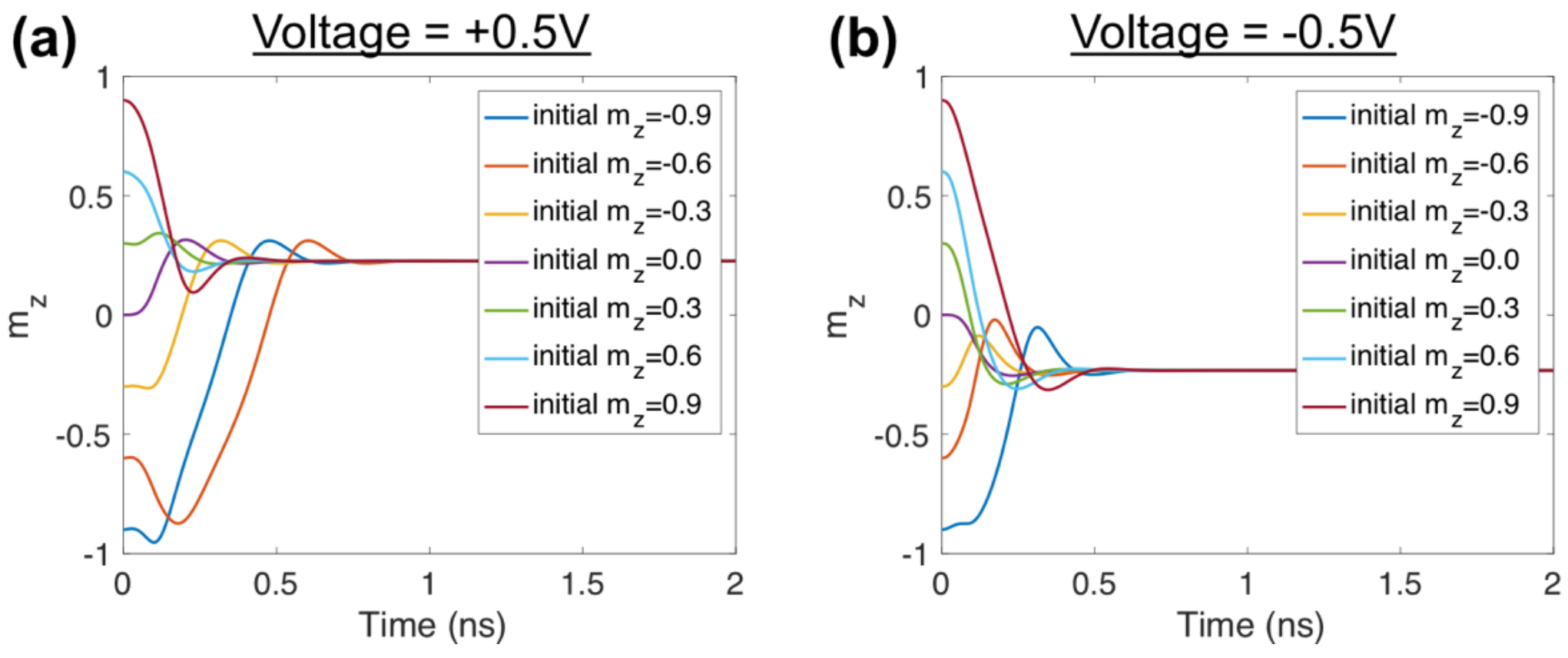

**Supplementary Figure S2.** Temporal change of volume-averaged perpendicular magnetization $m_z$ for different initial states for +0.5V and −0.5V applied voltages using the macrospin model.

## Supplementary Note 3. Switching speed

Supplementary Fig. S3 shows macrospin model results for the perpendicular component of magnetization $m_z$ is plotted as a function of time. The simulated structure is shown in Fig. 1a in the main text. Each subplot in Supplementary Fig. S3 has a different biaxial strain level (shown in the plot legends), while the current density is varied from $4 \times 10^7$ A/cm$^2$ to $1 \times 10^8$ A/cm$^2$ in each subplot. The dashed vertical lines in Supplementary Fig. S3 (a)-(d) indicate the fastest switching case in each case studied. In every case, the fastest switching speed occurs when the applied current reaches the maximum value of $1 \times 10^8$ A/cm$^2$. As can be seen, increasing strain does not increase switching speed, but it does increase the amplitude of the switching, i.e. the amplitude of $|m_z|$ at t

= 1 ns. The fastest switching speed of the strain-mediated SOT switching method is ~0.1 ns corresponding to an optimistic writing speed of ~10 GHz.

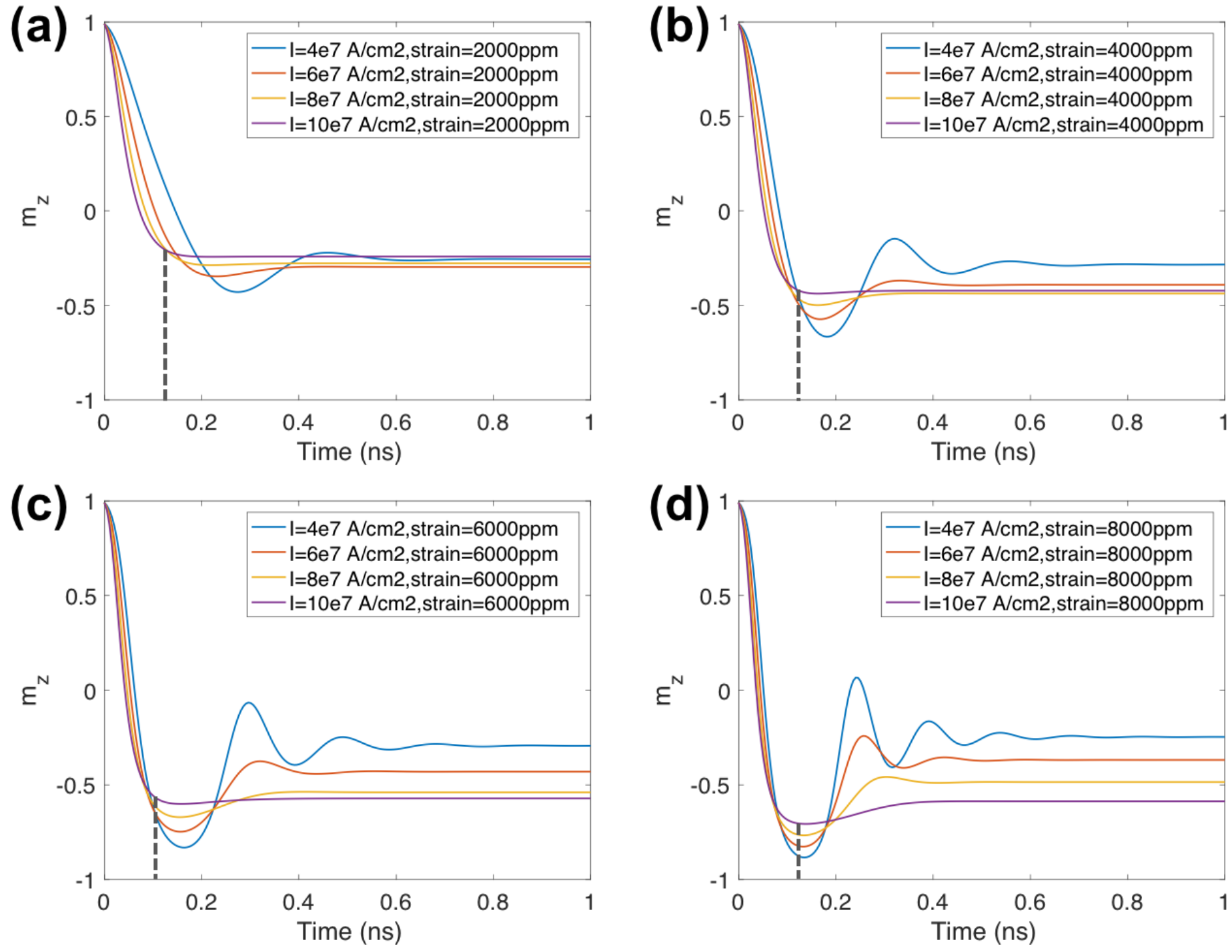

**Supplementary Figure S3**. Macrospin results of varying current and voltage for fixed angle θ = 45°. The dashed vertical lines indicate the fastest switching case for each plot.

**Supplementary Note 4. Mirror symmetry analysis of symmetry breaking**

Supplementary Fig. S4-1 provides an explanation of the mirror symmetry rules used in this manuscript. As shown in Supplementary Fig. S4-1(a), for ordinary vectors such as velocity, any component that is perpendicular to the mirror reverse their direction in the mirror reflection. However, any vector component parallel to the mirror remains in their original direction after the mirror reflection. The opposite is true for a pseudovector (or axial vector) such as magnetic field or magnetization. As shown in Supplementary Fig. S4-1(b),[1] for these pseudovectors, any component perpendicular to the mirror remains in its original direction in the mirror reflection. However, any pseudovector component parallel to the mirror reverses its direction in the mirror reflection.

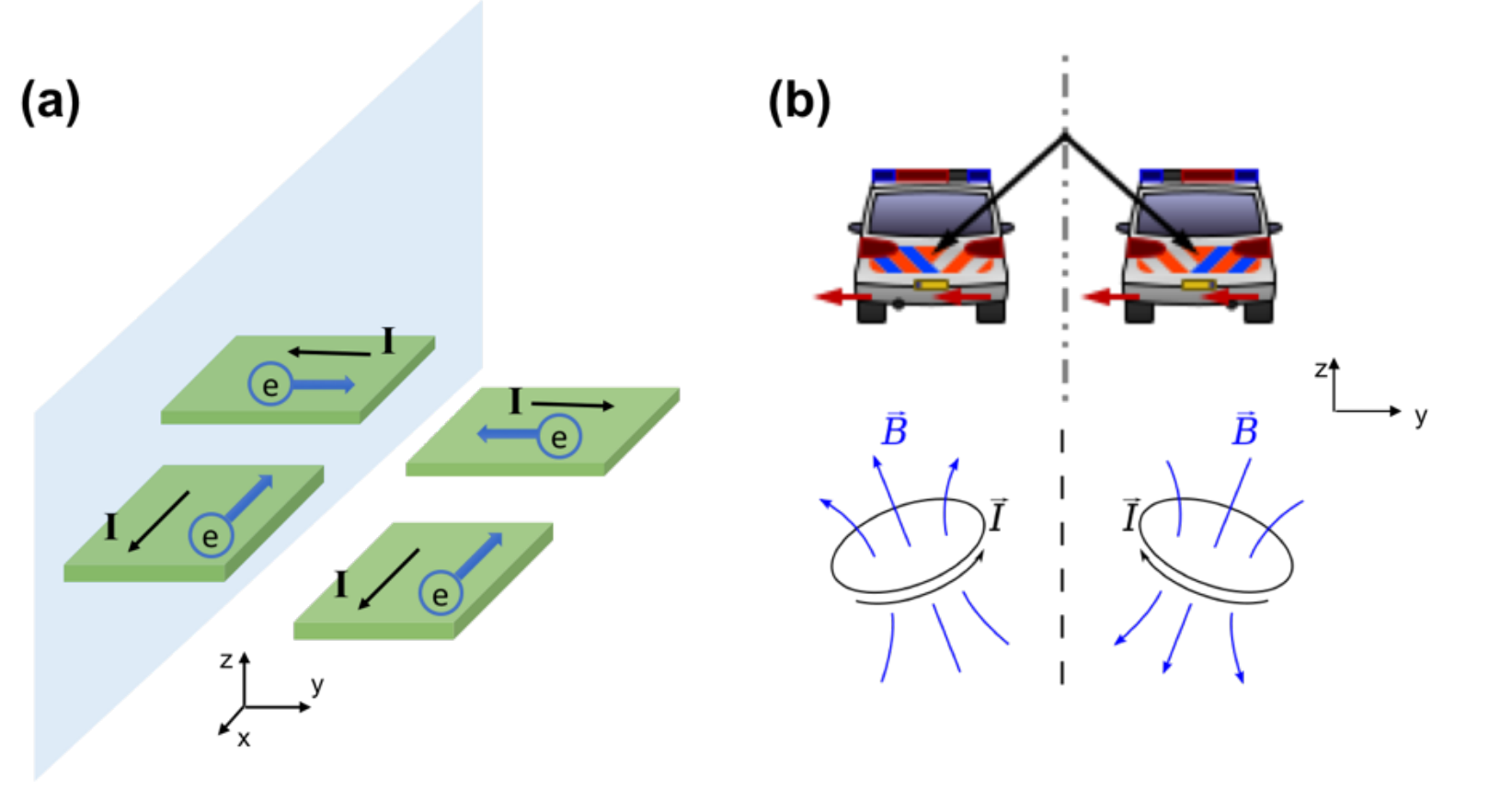

**Supplementary Figure S4-1.** Mirror symmetry rules. (a) Ordinary vector (or polar vector), such as velocity. (b) Pseudovector (or axial vector), such as magnetization, magnetic field, etc.

Supplementary Fig. S4-2 shows the mirror symmetry analysis for two situations: (a) applying pure SOT current; (b) applying SOT current combined with a magnetic bias field. Both scenarios consider a ferromagnetic layer with PMA (the green sheet). The blue parallelograms represent the mirror in the *xz* plane. The SOT current is applied along the –*x* axis, therefore the induced electron polarization is along –*y* axis, i.e. $\hat{\boldsymbol{\sigma}} = -\hat{y}$. Therefore, the damping-like SOT field is $\boldsymbol{H}_{SOT-DL} \propto \boldsymbol{m}\times\hat{\boldsymbol{\sigma}} = \hat{y}\times\boldsymbol{m}$. The current remains in its original direction in the mirror image because the velocity of an electron is ordinary vector. There are two pseudovectors in Supplementary Fig. S4-2(a): magnetization and the SOT field. Since both are parallel to the mirror plane, the two pseudovectors reverse their directions in the mirror image. The inputs in the real world and its mirror image remains the same, with both having the SOT current applied along –*x* axis. However, the two configurations produce opposite magnetization states. In other words, both "up" and "down" magnetization configurations exist with SOT current applied along –*x* axis. There is no preferable perpendicular direction, and the symmetry remains unbroken. Therefore, non-deterministic switching occurs. In Supplementary Fig. S4-2(b), a bias field along –*x* axis is added to the system. Since the bias field is also an axial vector and is parallel to the mirror plane, it reverses its direction in the mirror image. The two configurations in the real world and mirror image now have different inputs: the currents are the same but the bias fields are in opposite directions. The bias field added to the SOT system breaks the symmetry. A unique $m_z$ results based

on the $\boldsymbol{H}_{Bias}$ and current direction applied. In Supplementary Fig. S4-2(b), the bias field along –*x* axis prefers "up" while the bias field along +*x* axis prefers "down". This produces deterministic switching.

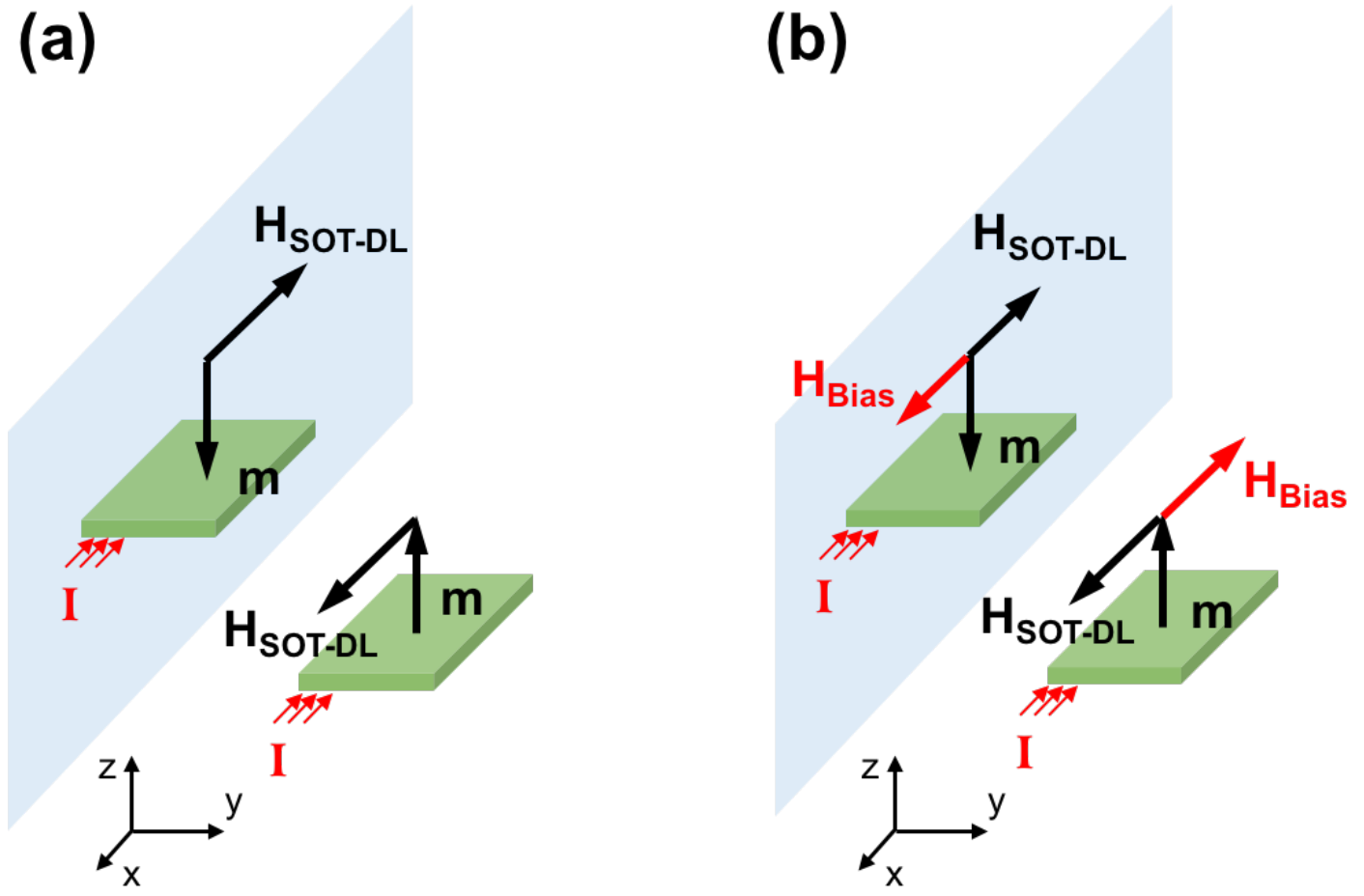

**Supplementary Figure S4-2.** Mirror symmetry analysis with respect to *xz* plane. (a) Pure SOT system. (b) Adding magnetic bias field to SOT system breaks in-plane symmetry.

Supplementary Fig. S4-3 shows mirror symmetry analysis with respect to *xz* plane for strain-mediated systems. While the field-like SOT term is absent in the simulations for simplicity, both damping-like and field-like spin-orbit torques are considered here in the mirror symmetry analysis to provide a more complete understanding. Similar to Supplementary Fig. S2, the two scenarios considered are a ferromagnetic layer with PMA (the green sheet) and the blue parallelograms represent the mirror in the *xz* plane. Also, the electron polarization is along –y axis, i.e. $\hat{\boldsymbol{\sigma}} = -\hat{y}$. Therefore, the damping-like SOT field is $\boldsymbol{H}_{SOT-DL} \propto \boldsymbol{m}\times\hat{\boldsymbol{\sigma}} = \hat{y}\times\boldsymbol{m}$ and the field-like SOT field is $\boldsymbol{H}_{SOT-FL} \propto \hat{\boldsymbol{\sigma}} = -\hat{y}$.

In Supplementary Fig. S4-3(a), a uniaxial tensile strain along *x* axis is applied to the SOT system such that the strain and applied current are parallel to each other. This introduces a magnetoelastic field $\boldsymbol{H}_{ME} \propto m_y\hat{\boldsymbol{x}}$ (see Eq. 2 in main text of the paper). Because the strain is uniaxial, a bidirectional arrow along the *x* axis is drawn to represent $\boldsymbol{H}_{ME}$. Following the mirror symmetry rule of an axial vector, the two parts (yellow and red) of $\boldsymbol{H}_{ME}$ both reverse their directions after mirror reflection. However, overall $\boldsymbol{H}_{ME}$ still looks the same since the bidirectional arrow sits along the $x$ axis in both configurations. The magnetization and damping-like field reverse their directions as they are pseudovectors and are parallel to the mirror. The field-like field remains the direction as it is perpendicular to the mirror. In Supplementary Fig. S4-3(a), both configurations in the real world and mirror image have the same inputs: current along $-x$ direction and $\boldsymbol{H}_{ME}$ along $x$ direction. However, they have opposite magnetization states. In other words, both "up" and "down" magnetization configurations are allowed when the strain and current are parallel. This results in an absence of a preferable perpendicular direction, and the symmetry is unbroken. Therefore, deterministic switching cannot occur. One can easily obtain the same conclusion, i.e., no deterministic switching happens when the strain is perpendicular to the current.

In Supplementary Fig. S4-3(b), the $\boldsymbol{H}_{ME}$ is canted from the current. Following the mirror symmetry rule of pseudovectors, the two parts (yellow and red) of $\boldsymbol{H}_{ME}$ both change directions after mirror reflection transformation. $\boldsymbol{H}_{ME}$ arrows change relative orientation in its mirror image. All other vectors have the same directions with Supplementary Fig. S4-3(a), thus the mirror reflections of them are not repeated here. In Supplementary Fig. S4-3(b), the configurations in the "real world" and the "mirror world" have different inputs: same current along $-x$ direction but different $\boldsymbol{H}_{ME}$. The symmetry is broken when the strain is canted from the current. A unique $m_z$ (either up or down) arises based on $\boldsymbol{H}_{ME}$ and current direction, and deterministic switching is allowed.

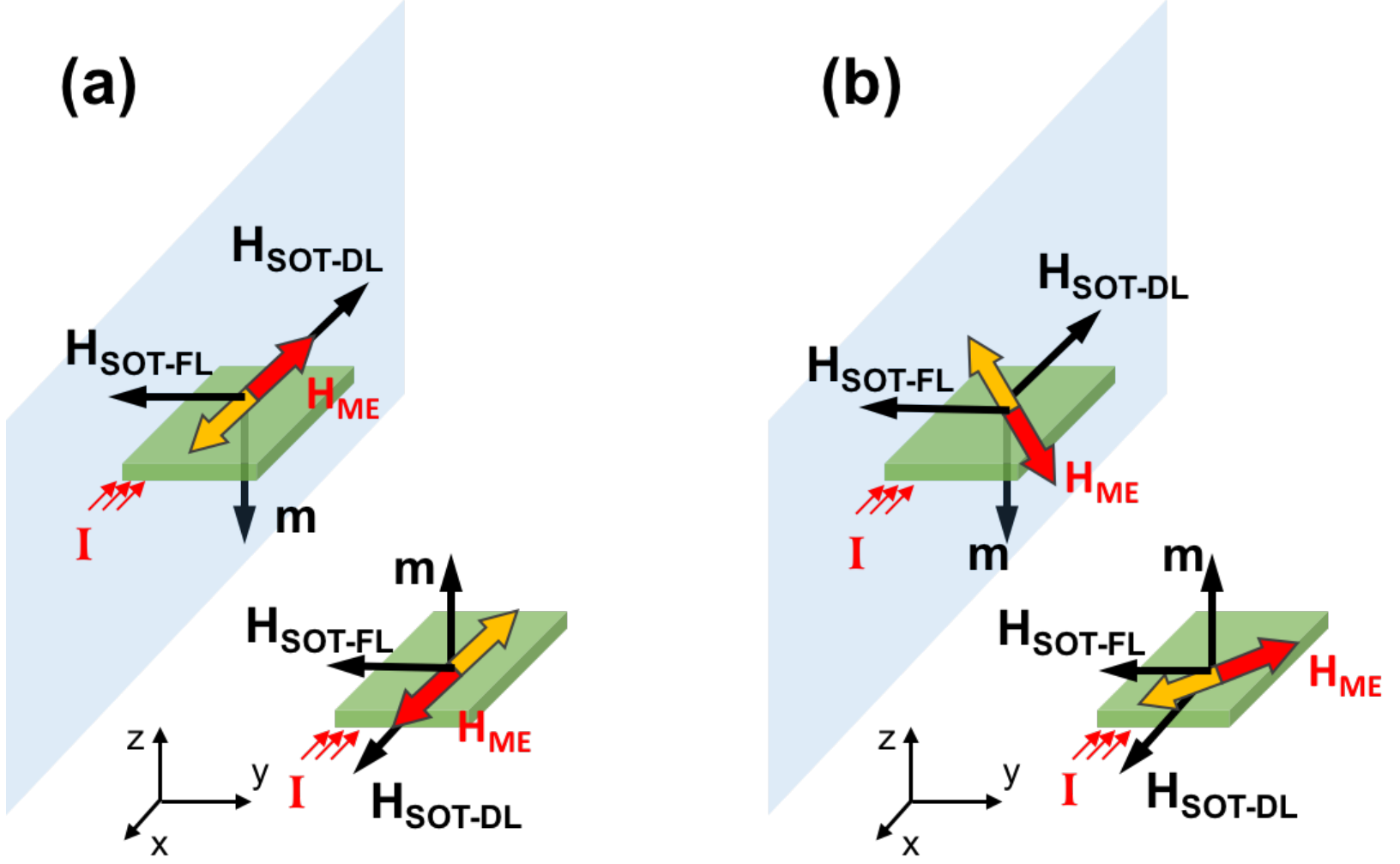

**Supplementary Figure S4-3.** Mirror symmetry analysis with respect to *xz* plane for strain-mediated SOT system with both damping-like and field-like SOT fields considered. The green sheet represents the ferromagnetic layer with PMA, while the blue parallelogram represents the mirror in *xz* plane. (a) Mirror symmetry analysis when magnetoelastic fields $\boldsymbol{H}_{ME}$ is parallel to the current. The 'real world' and the 'mirror world' have the same inputs (i.e., current and $\boldsymbol{H}_{ME}$) but opposite magnetization states. Therefore, both magnetization states exist given the same inputs, and the symmetry is not broken. (b) Mirror symmetry analysis when $\boldsymbol{H}_{ME}$ is canted from the current. The $\boldsymbol{H}_{ME}$ changes orientation in its mirror image. Therefore, the 'real world' and 'mirror world' have different inputs. A unique magnetization is chosen for each $\boldsymbol{H}_{ME}$ orientation and the lateral symmetry is broken.

**Supplementary Note 5. Mathematical explanation for deterministic switching**

As a complementary explanation to mirror symmetry analysis, we provide a mathematical approach to examine the stochastic nature of the switching process for different physical inputs (e.g., current and strain). The precessional magnetic dynamics are governed by the Landau-Lifshitz-Gilbert (LLG) equation with SOT terms:[2-5]

$$\frac{\partial \boldsymbol{m}}{\partial t} = -\mu_0 \gamma (\boldsymbol{m} \times \boldsymbol{H}_{eff}) + \alpha \left( \boldsymbol{m} \times \frac{\partial \boldsymbol{m}}{\partial t} \right) - \frac{\gamma \hbar}{2e} \frac{J_C}{M_S t_F} \xi_{DL} \boldsymbol{m} \times (\boldsymbol{m} \times \widehat{\boldsymbol{\sigma}}) \qquad \text{(Eq. S1)}$$

Refer to Eq. (1) in the main text for parameter explanations. As shown in last section S4, the field-like torque does not influence the symmetry-breaking. Therefore, in this section, the field-like torque is neglected for simplification. The stable magnetization state is determined by solving the LLG equation with $\boldsymbol{\tau}_{tot} = 0$, then the equilibrium equation becomes:

$$\boldsymbol{\tau}_{tot} = 0 = -\mu_0\gamma\left(\boldsymbol{m}\times\boldsymbol{H}_{eff}\right) - \frac{\gamma\hbar}{2e}\frac{J_C}{M_S t_F}\xi_{DL}\boldsymbol{m}\times(\boldsymbol{m}\times\hat{\boldsymbol{\sigma}}) \qquad \text{(Eq. S2)}$$

$\boldsymbol{H}_{eff}$ is the effective field and there are two dominating terms: $\boldsymbol{H}_{eff} \approx \boldsymbol{H}_{PMA} + \boldsymbol{H}_{ME}$, where $\boldsymbol{H}_{PMA}$ is the effective PMA field, and $\boldsymbol{H}_{ME}$ is the magnetoelastic field. The PMA field can be generalized using the following approach considering a phenomenological PMA coefficient:[6,7]

$$\boldsymbol{H}_{PMA} = -\frac{2}{\mu_0 M_S}K_{PMA}m_z\hat{\boldsymbol{z}} \qquad \text{(Eq. S3)}$$

For a certain magnetic materials and geometries, the PMA coefficient is a constant. Refer to Eq. (2) in the main text for magnetoelastic field $\boldsymbol{H}_{ME}$. For the positive magnetostrictive material CoFeB, both $B_1$ and $B_2$ are negative constants. Consider a simple case: $\varepsilon_{yy} > 0$, all other strains are zeros. Then the magnetoelastic field can be simplified as:

$$\boldsymbol{H}_{ME} = -\frac{2}{\mu_0 M_S}B_1\varepsilon_{yy}m_y\hat{\boldsymbol{y}} \qquad \text{( Eq. S5)}$$

Plugging Eq. S4 and Eq. S5 into Eq. S2, we get the reduced equilibrium equation:

$$\boldsymbol{\tau}_{tot} = 0 = -\boldsymbol{m}\times\left(\mathrm{A}m_z\hat{\boldsymbol{z}} + \mathrm{B}m_y\hat{\boldsymbol{y}}\right) - \mathrm{C}[\boldsymbol{m}\times(\boldsymbol{m}\times\hat{\boldsymbol{\sigma}})] \qquad \text{( Eq. S6)}$$

where $A = -\frac{2\gamma}{M_S}K_{PMA}$, $B = -\frac{2\gamma}{M_S}B_1\varepsilon_{yy}$, $C = \frac{\gamma\hbar}{2e}\frac{J_C}{M_S t_F}\xi_{DL}$ are three positive constants related to PMA, strain and current density, respectively. Consider a general in-plane polarization $\hat{\boldsymbol{\sigma}} = (\sigma_1, \sigma_2, 0)$. The equilibrium equation Eq. S6 can be expressed as:

$$\boldsymbol{\tau}_{tot} = \begin{pmatrix} -(A-B)m_y m_z + C(m_y^2\sigma_1 + m_z^2\sigma_1 - m_x m_y\sigma_2) \\ A m_x m_z + C(m_x^2\sigma_2 + m_z^2\sigma_2 - m_x m_y\sigma_1) \\ -B m_x m_y - C(m_x\sigma_1 + m_y\sigma_2)m_z \end{pmatrix} = \begin{pmatrix} 0 \\ 0 \\ 0 \end{pmatrix} \qquad \text{( Eq. S7)}$$

Under the constrain that:

$$m_1^2 + m_2^2 + m_3^2 = 1 \quad \text{( Eq. S8)}$$

The magnetization state $\boldsymbol{m} = (m_x, m_y, m_z)$ represents the unknown to be solved. The governing equations Eq. S7 and Eq. S8 are homogeneous, which means if $(m_{10}, m_{20}, m_{30})$ is a solution, then $(-m_{10}, -m_{20}, -m_{30})$ is also a solution. In other words, there can be more than one solution to the equilibrium equation. However, not every solution represents a stable magnetization state. As shown in Supplementary Fig. S5, the equilibrium state can be classified as either stable or unstable. The two types can be differentiated by checking the divergence of total torque $\nabla \cdot \boldsymbol{\tau}_{tot}$. As shown in Supplementary Fig. S5 (a) and (b), $\nabla \cdot \boldsymbol{\tau}_{tot} < 0$ and $\nabla \cdot \boldsymbol{\tau}_{tot} > 0$ indicate stable equilibrium state and unstable equilibrium state, respectively. The divergence for Eq. S7 is:

$$\nabla \cdot \boldsymbol{\tau}_{tot} = -2C(m_x\sigma_1 + m_y\sigma_2) \quad \text{( Eq. S9)}$$

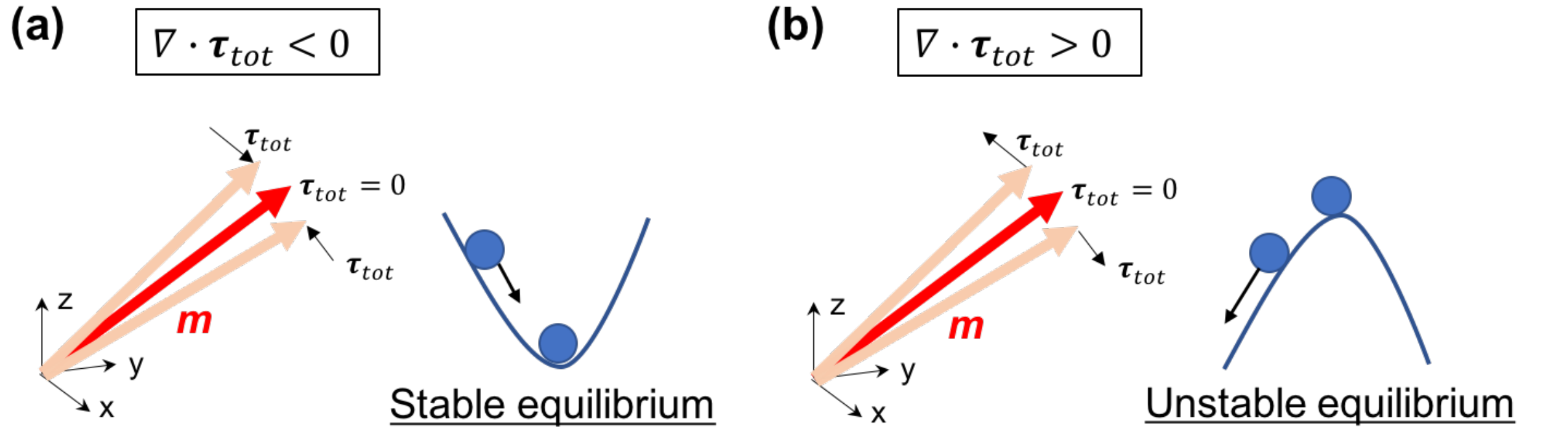

**Supplementary Figure S5.** Stable and unstable equilibriums. (a) Negative divergence of $\boldsymbol{\tau}_{tot}$ indicates stable equilibrium. (b) Positive divergence of $\boldsymbol{\tau}_{tot}$ indicates unstable equilibrium

Case 1: consider only a SOT current is applied. For this case the equilibrium equation becomes:

$$\boldsymbol{\tau}_{tot} = \begin{pmatrix} -Am_ym_z + C(m_y^2\sigma_1 + m_z^2\sigma_1 - m_xm_y\sigma_2) \\ Am_xm_z + C(m_x^2\sigma_2 + m_z^2\sigma_2 - m_xm_y\sigma_1) \\ -C(m_x\sigma_1 + m_y\sigma_2)m_z \end{pmatrix} = \begin{pmatrix} 0 \\ 0 \\ 0 \end{pmatrix} \quad \text{( Eq. S10)}$$

(i) If $m_z = 0$, one can solve for $m_x$ and $m_y$ by:

$$\begin{cases} m_y^2\sigma_1 - m_x m_y \sigma_2 = 0 \\ m_x^2\sigma_2 - m_x m_y \sigma_1 = 0 \\ m_x^2 + m_y^2 = 1 \\ \sigma_1^2 + \sigma_2^2 = 1 \end{cases} \qquad \text{( Eq. S11)}$$

The solution is simply $m_x = \sigma_1$, $m_y = \sigma_2$, or $m_x = -\sigma_1$, $m_y = -\sigma_2$, i.e. the magnetization follows the direction of electron polarization.

(ii) If $m_z \neq 0$, then there is a paired solution of pointing either up or down. Also from the 3rd line of Eq. S10, we get:

$$m_x\sigma_1 + m_y\sigma_2 = 0 \qquad \text{( Eq. S12)}$$

Therefore, the divergence of total torque is always zero:

$$\nabla \cdot \boldsymbol{\tau}_{tot} = -2C(m_x\sigma_1 + m_y\sigma_2) = 0 \qquad \text{( Eq. S13)}$$

This means that the two states (up or down) in each paired solution are both neutral equilibrium states. In conclusion, if SOT is the only input to the system, either the magnetization is forced to follow the electron polarization in-plane, or there are symmetric neutral equilibrium states out-of-plane, and deterministic switching is not produced.

Case 2: consider the current is applied along –*y* axis, which is parallel to the strain. The electron polarization has to be perpendicular to the strain (i.e., *x* axis), so let $\hat{\boldsymbol{\sigma}} = (1,0,0)$. The equilibrium equation becomes:

$$\boldsymbol{\tau}_{tot} = \begin{pmatrix} -(A-B)m_y m_z + C(m_y^2 + m_z^2) \\ A m_x m_z - C m_x m_y \\ -B m_x m_y - C m_x m_z \end{pmatrix} = \begin{pmatrix} 0 \\ 0 \\ 0 \end{pmatrix} \qquad \text{( Eq. S14)}$$

(i) If $m_x \neq 0$, then a closed form solution can be obtained by solving the 2nd and 3rd line of Eq. S14. The solution is $\boldsymbol{m}$ = (±1,0,0) with magnetization in-plane along the x axis.

(ii) If $m_x = 0$, symmetric solutions arise to Eq. S14. However, the divergence of total torque is always zero as:

$$\nabla \cdot \boldsymbol{\tau}_{tot} = -2Cm_x = 0 \qquad \text{( Eq. S15)}$$

This means the out-of-plane solution, if there is any, is always a neutral equilibrium state. In conclusion, if the system has strain and current inputs that are parallel to each other, then either magnetization follows the electron polarization in-plane, or there are symmetric equilibrium states out-of-plane. Therefore, deterministic switching is not produced.

Case 3: consider the situation when the current is canted from the strain. For example, let $\hat{\boldsymbol{\sigma}} = (1,1,0)/\sqrt{2}$. The equilibrium equation becomes:

$$\boldsymbol{\tau}_{tot} = \begin{pmatrix} -(A-B)m_y m_z + C(m_y^2 + m_z^2 - m_x m_y)/\sqrt{2} \\ A m_x m_z + C(m_x^2 + m_z^2 - m_x m_y)/\sqrt{2} \\ -B m_x m_y - C(m_x + m_y)m_z/\sqrt{2} \end{pmatrix} = \begin{pmatrix} 0 \\ 0 \\ 0 \end{pmatrix} \qquad \text{( Eq. S16)}$$

The divergence of the total torque is:

$$\nabla \cdot \boldsymbol{\tau}_{tot} = -\sqrt{2}C(m_x + m_y) \qquad \text{( Eq. S17)}$$

Obviously there is an absence of in-plane solutions because: if one let $m_z = 0$, then there is only trivial solution $\boldsymbol{m}$ = (0,0,0) to Eq. S16. However, this does not satisfy Eq. S8. Therefore, no in-plane solution is possible for Eq. S16. In other words, the equilibrium state is always out-of-plane. Inferred from Eq. S17, in the paired solutions $\pm(m_{10}, m_{20}, m_{30})$, there is always one stable state and one unstable state, i.e. a specific direction (either up or down) is stable. Therefore, deterministic switching is produced.

Using this method, one can also arrive at the conclusion that applying bias field leads to deterministic switching. In conclusion, checking the divergence of the total torque is an equivalent method to mirror symmetry analysis, and both methods predict whether deterministic switching is possible.

**Supplementary Note 6. Criterion for switching phase diagrams**

As shown in Supplementary Fig. S6, there are four types of magnetization state after application of voltage/current. A criterion is provided to differentiate those behaviors in the macrospin code. The magnetization is initialized as 'up' (i.e., $m_z = +1$), then the temporal evolution of magnetization is performed for each combination of applied current, strain and angle between them. The perpendicular magnetization $m_z$ after 2 ns voltage/current application is investigated. There are four types of possible magnetization states: (I) magnetization reverses its direction and stabilizes in the opposite phase; (II) magnetization experiences 90° switching and stabilizes in-plane; (III) magnetization experiences less than 90° switching and stabilizes in the original direction; or (IV) magnetization continuously oscillates across in-plane without stabilizing in any preferred direction. If the variation of $m_z$ between t = 2 ~ 3 ns is less than 5%, then the magnetization at 2ns is considered to be stabilized. Under the stabilized state, $m_z > 0.05$, $|m_z| < 0.05$ and $m_z < -0.05$ corresponds to type I, II, and III respectively. If the variation of $m_z$ is greater than 5%, and $m_z$ keeps changing its sign in the last 1ns, then it is classified as type IV. It is worth noting that the existence of type III does not conflict with the conclusion in S5 that in-plane equilibrium states do not exist for this combination strain/current. Mathematically there is absence of an exact "in-plane" solution (i.e., $m_z = 0$), however, any solution with $-0.05 < m_z < 0.05$ produces an absence of preferable perpendicular directions under thermal fluctuation. Therefore, we classify those solutions as "in-plane" solution.

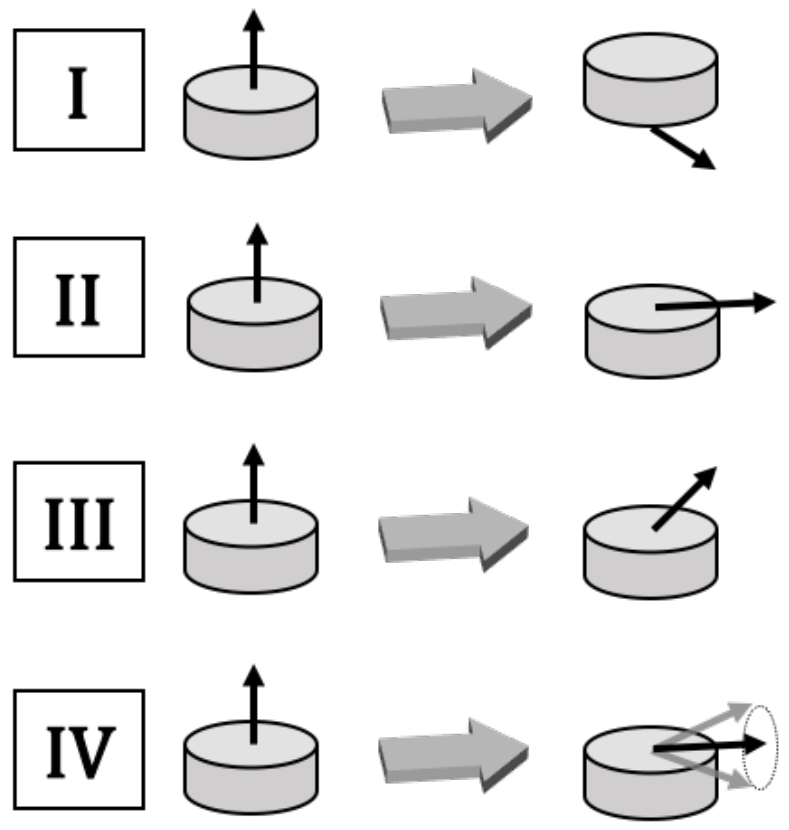

**Supplementary Figure S6.** Four types of magnetization state after application of strain/current.

**Supplementary References:**

1. Wikipedia contributors. "Pseudovector." *Wikipedia, The Free Encyclopedia*. Wikipedia, The Free Encyclopedia, 12 Oct. 2017. Web.
2. Fukami, S., et al. "A spin–orbit torque switching scheme with collinear magnetic easy axis and current configuration." *Nature nanotechnology* 11.7 (2016): 621-625.
3. Macrospin modeling of sub-ns pulse switching of perpendicularly magnetized free layer via spin-orbit torques for cryogenic memory applications.
4. Finocchio, G., et al. "Switching of a single ferromagnetic layer driven by spin Hall effect." *Applied Physics Letters* 102.21 (2013): 212410.
5. Li, Peng, et al. "Spin-orbit torque-assisted switching in magnetic insulator thin films with perpendicular magnetic anisotropy." *Nature communications* 7 (2016).
6. Wang, Qianchang, et al. "Strain-mediated 180° switching in CoFeB and Terfenol-D nanodots with perpendicular magnetic anisotropy." *Applied Physics Letters* 110.10 (2017): 102903.
7. Li, Xu, et al. "Strain-mediated 180 perpendicular magnetization switching of a single domain multiferroic structure." *Journal of Applied Physics* 118.1 (2015): 014101.